\documentclass[%
 aip,
 amsmath,amssymb,
 reprint,%
]{revtex4-1}

\usepackage{xr}
\usepackage{float}
\usepackage[colorlinks=true]{hyperref}

\usepackage{soul}
\usepackage{color}

\usepackage{graphicx}% Include figure files
\usepackage{dcolumn}% Align table columns on decimal point
\usepackage{bm}% bold math
\renewcommand{\textcolor}[2]{#2}
\renewcommand{\color}[1]{}

\usepackage[utf8]{inputenc}
\usepackage[T1]{fontenc}
\usepackage{mathptmx}
\usepackage{etoolbox}

\usepackage{mathrsfs} 

\makeatletter
\def\@email#1#2{%
 \endgroup
 \patchcmd{\titleblock@produce}
  {\frontmatter@RRAPformat}
  {\frontmatter@RRAPformat{\produce@RRAP{*#1\href{mailto:#2}{#2}}}\frontmatter@RRAPformat}
  {}{}
}%
\makeatother
\begin{document}

\preprint{AIP/123-QED}

\title[Estimation and measurement of PSFs for SKPM]{
    Beyond the blur: using experimentally determined point spread functions to improve scanning Kelvin probe imaging}
\author{Isaac C.D. Lenton*}
 \email{isaac.lenton@ist.ac.at}
\author{Felix Pertl}
\author{Lubuna Shafeek}
\author{Scott R. Waitukaitis}
\affiliation{ 
$^1$Institute of Science and Technology Austria, Am Campus 1, 3400 Klosterneuburg, Austria
}

\date{\today}% It is always \today, today,

\begin{abstract}
Scanning Kelvin probe microscopy (SKPM) is a powerful technique for
investigating the electrostatic properties of material surfaces,
enabling the imaging of variations in work function, topology, surface charge
density, or combinations thereof.
Regardless of the underlying signal source,
SKPM results in a voltage image which is spatially
distorted due to the finite size of the probe, long-range
electrostatic interactions, mechanical and electrical
noise, and the finite response time of the electronics.
In order to recover the underlying signal, it is necessary to deconvolve
the measurement with an appropriate point spread function (PSF) that accounts
the aforementioned distortions, but determining this PSF is difficult.
Here we describe how such PSFs can be
determined experimentally, and show how they can be used to recover
the underlying information of interest.
We first consider the physical principles that enable SKPM, and discuss how these affect the system PSF. We then show how one can experimentally measure PSFs by looking at
well defined features, and that these compare well to simulated PSFs, provided scans are
performed extremely slowly and carefully.
Next, we work at realistic scan speeds, and show that the idealised PSFs fail to
capture temporal distortions in the scan direction.
While simulating PSFs for these situations would be quite challenging,
we show that measuring PSFs with similar scan \textcolor{blue}{conditions} works well.
Our approach clarifies the basic principles of and inherent challenges to
SKPM measurements, and gives practical methods to improve results.
\\[10px]
%\textbf{This is a pre-print.  Please check for corrections/modifications when the article is published.}
\textbf{Pre-print of:}
J. Appl. Phys. 136, 045305 (2024)
\url{https://doi.org/10.1063/5.0215151}
%\textbf{This is a pre-print.  The
%following article has been submitted to
%Journal of Applied Physics.
%Please check for corrections/modifications when the article is published.}
\\Copyright \copyright~2024.
This manuscript version is made available under the CC-BY-NC-ND 4.0\\
license \url{http://creativecommons.org/licenses/by-nc-nd/4.0/}.
\end{abstract}

\maketitle

\section{Introduction}
\begin{figure}[h!]
\includegraphics[width=3.37in]{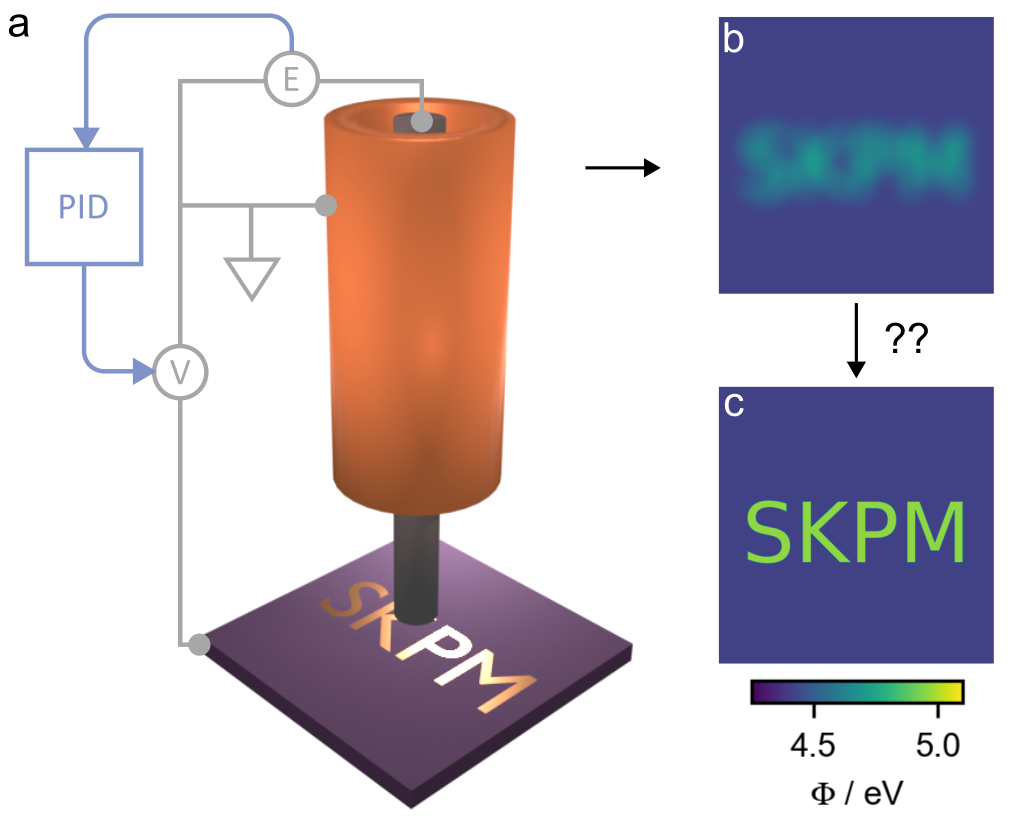}
\caption{\label{fig:figure1}
\textbf{Principle of scanning Kelvin probe microscopy (SKPM).}
\textbf{a}~SKPM entails scanning a conductive probe above a sample to estimate the local  surface potential. The cylindrical metal probe is vibrated vertically, while a lock-in amplifier and PID use feedback adjust the backing potential, $V$, so that the current in the probe is minimised.  
\textbf{b}~Voltage maps measured by SKPM are spatially distorted, due to the probe's finite size, long-range electrostatics, and the temporal response of the electronics. \textbf{c}~This paper is about extracting the true underlying surface potential from the measured signal, which we show is achievable by experimentally measuring the system's PSF and using this to deconvolve raw measurements.}
\end{figure}

SKPM  enables imaging of the ``invisible'' 
electrostatic properties of surfaces at the mesoscale (typically $>$100\,$\text{\textmu m}$). 
By scanning a vibrating conductive probe above a surface and measuring/regulating the current induced within it (Fig.~\ref{fig:figure1}\textbf{a}), SKPM extracts information connected to the local electric potential\citep{Zisman1932Jul, Craig1970Feb}.
This gives insight into various processes such as variations in work function or surface chemistry\citep{Nazarov2019Aug, Nazarov2012},
charge\citep{Baytekin2011Jul, Bai2021Jun},
biological double layers\citep{Hackl2022Nov, Martinsen2023}, \textit{etc.}
Though not identical, the operating principle behind SKPM is  related to Kelvin probe force microscopy (KPFM)\citep{Nonnenmacher1991Jun}, where measurement/regulation of the forces acting on a vibrating atomic force microscope tip 
permits similar imaging at the nanoscale\citep{Glatzel2022Feb, Melitz2011Jan, Axt2018Jun}. 
\textcolor{blue}{
One major advantage of SKPM over KPFM is the use of significantly larger probes
enabling rapid scans over significantly larger areas (a discussion on
the differences/similarities between the two techniques is included in
the Supplementary Material, SM).}
%\footnote{\textit{e.g.}, $10^8$~\text{\textmu m}$^2$ \textit{versus}
%$10^4$~\text{\textmu m}$^2$ is typical on several commercial instruments.}.}

A fundamental problem in both SKPM and KPFM is properly interpreting and analyzing the measured signal. Regardless of the signal source (\textit{e.g.}~work function variations or charge), both techniques
produce voltage images. Moreover, these images are spatially distorted due
to the finite size of the probe,
long-range electrostatic
interactions\citep{Cohen2013Jun, Machleidt2009Jun}, mechanical and electrical noise\citep{Ren2023May},
and the temporal response of the electronics\citep{Checa2023Nov, Ziegler2013Jul, Craig1970Feb}.
When these distortions are deterministic and linear, they can be characterised
by a PSF, which describes the image an infinitesimally small point in the absence of noise would produce
when measured. Effectively, the measured signal is the convolution of
the underlying signal with the PSF, and 
the underlying signal can be recovered by deconvolution\citep{Cohen2013Jun, Machleidt2009Jun}.
Therefore, a key component in interpreting/analyzing SKPM (or KPFM) data relies on being able to determine the PSF that characterises the measurement process.
%\textcolor{blue}{This is especially true for fast scans performed
%with large probes, where properly characterising and compensating
%for these distortions is important for achieving the best possible resolution.}

Toward addressing the problem, one can draw inspiration from optical microscopy. Like SKPM, optical images suffer from distortions, due to diffraction, optical imperfections, finite pixel sizes \textit{etc.}
To correct for these, a practical solution is to image the pattern created by a point-like emitter, \textit{e.g.}~a small fluorescent particle\citep{Cole2011Dec}.
When this target is significantly smaller than the system resolution, the image rendered is approximately the PSF. 
%This is because it is not trivial
%to create well characterised calibration targets
%with a large enough signal to produce a PSF with sufficiently low noise.
%\hl{For instance, in KPFM the probes are $\gtrsim$ 10 nm in size, which puts creating PSF targets beyond practical reach given they be significantly smaller (\textit{e.g.}~$\sim$1 nm). Naively, it should seem that this approach is more amenable to SKPM given the larger lengthscales involved, but to our knowledge this has not been demonstrated.}

Here, we show that the approach of experimentally measuring PSFs is not only viable, but effective and straightforwardly implemented in the case of SKPM.
\textcolor{blue}{Compared to KPFM, the larger probes used in SKPM allow scanning significantly larger areas at higher scan rates.
Consequently, for SKPM it is possible to experimentally measure the PSFs using relatively large calibration targets.}
We restrict ourselves to the situation where the underlying signal is due to differences in
material work functions on a planar surface, though in principle our ideas can be extended to other situations (\textit{e.g.}~variations in surface charge).
We use common clean-room techniques to pattern regions with work function differences, which we image to extract PSFs.
%\hl{We show that, despite the size advantage of SKPM over KPFM, care must be taken to overcome system noise, especially during deconvolution.
We find that when utilizing high scan speeds---which are necessary to probe
large \textcolor{blue}{and/or time dependent} features---the measured PSFs can differ significantly from
those that only account for the electrostatic interactions between the probe and
sample.
We show that this is due to incorporation of temporal information into the PSF---an issue that would be difficult to account for
analytically/computationally, but that is solved relatively easily in experiment.
\textcolor{blue}{
By simply measuring the PSF, we gain insight into the effect of
measurement speed, feedback parameters, and probe geometry that would
be difficult to predict or characterise independently.}
Our results outline a practical and easily implemented approach toward getting quantitative
information out of SKPM.

\section{Measurement Principal}
\label{sec:theory}

\subsection{Signal acquisition}

Figure~\ref{fig:figure1}\textbf{a} depicts an SKPM
probe positioned above a sample and the feedback system used to acquire an
estimate for the spatially varying surface potential of interest, $V_S\equiv V_S(x)$. The probe vibrates vertically at a fixed amplitude and frequency, while a potential ($V$) is applied to the electrode at or below the sample.
The current drawn to the probe due to the vibration is measured by an
electrometer ($E$) and further amplified by a lock-in amplifier which
extracts the current signal due to the probe vibration and rejects noise
at other frequencies.
The signal acquired by SKPM is the value of the voltage, $V$, required to minimise the current
induced in the probe as it is vibrated.
\textcolor{blue}{This is typically} accomplished by using feedback on the lock-in signal, \textit{e.g.}, sending it to proportional-integral-differential (PID) electronics and adjusting $V$ until the current amplitude is zero. %; \textcolor{blue}{or, measuring the current at a range of voltages and estimating the voltage from a fit to these measurements.}
In the na\"ive version of the analysis, the probe-sample system is assumed to form a capacitor, where the image charge drawn to the probe is given by
\begin{equation}
    Q(t) = C(h(t))(V - V_S).
    \label{eq:simple}
\end{equation}
Here, $C(h)$ is the capacitance, and
$h(t)$ is the time-varying height of the probe above the sample. By differentiating Eq.~\ref{eq:simple} with respect to time, setting equal to zero, and defining for $V_m\equiv V$ in this condition, we find
\begin{equation}
    0 = \frac{dQ}{dt} = \frac{dh}{dt}\frac{dC}{dh}(V_m - V_S).
\end{equation}
Hence, for finite derivatives $dh/dt$ and $dC/dh$, we see that this condition
is satisfied when
\begin{equation}
V_m(x)=V_S(x). 
\end{equation}
In other words, the simplest interpretation of raw SKPM data is that it is an exact, point-by-point copy of the surface potential of interest.

\begin{figure*}
\includegraphics[width=6.3in]{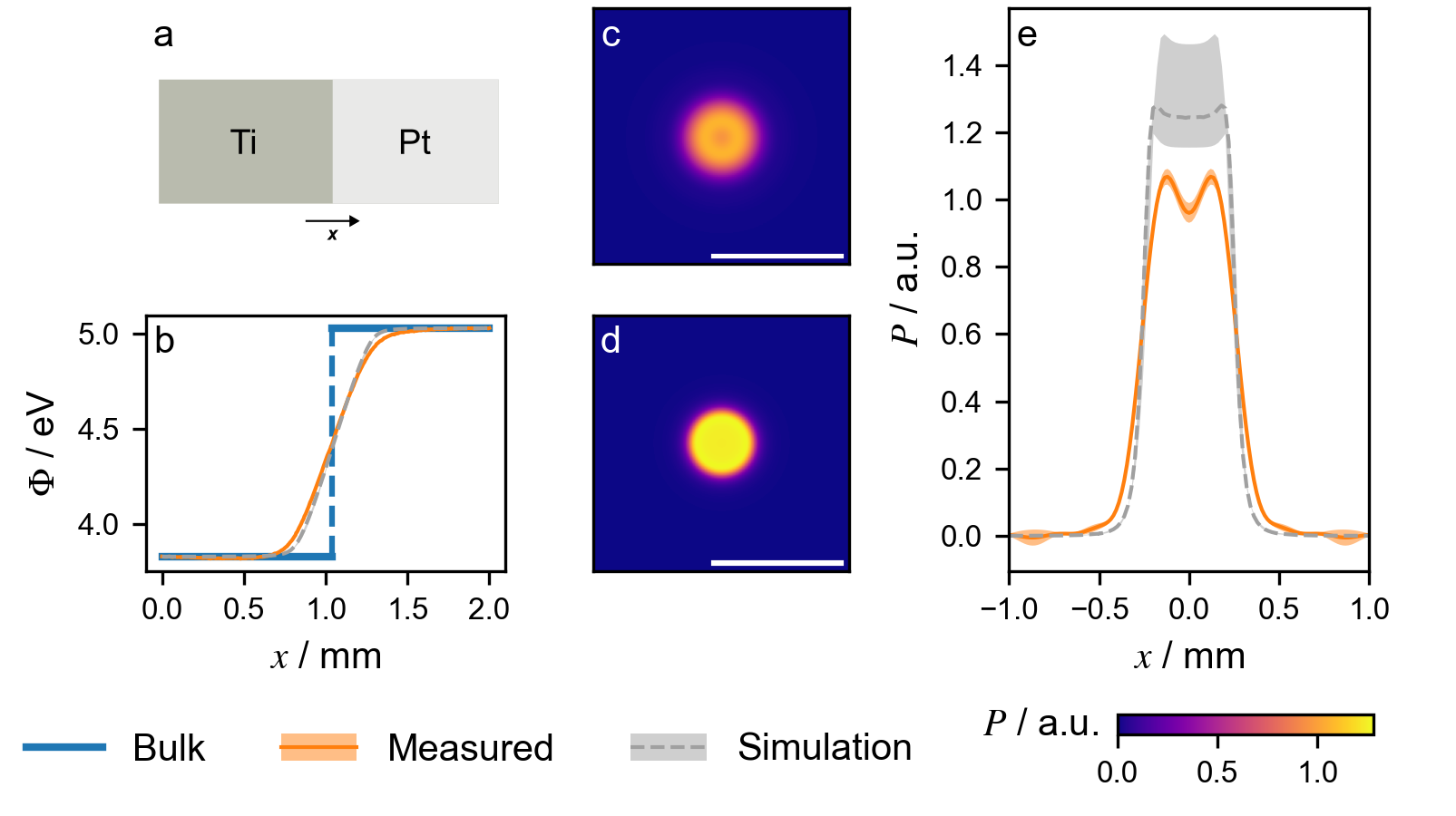}  % max 6.69
\caption{\label{fig:figure2}
\textbf{Comparison between a PSF estimated using an edge measurement and a simulated PSF.}
\textbf{a}~Scanning an edge created at the boundary between two metals with distinct work
functions produces \textbf{b}~an estimate for the system edge spread function (ESF)
along the scan direction.
\textbf{c}~Shows the system PSF estimated from the ESF measurement by assuming that
the probe is rotationally symmetric.
\textbf{d,e}~Simulations produce a PSF with a similar width and qualitatively similar
shape, but show non-negligible differences, \textcolor{blue}{even
for a range of different probe geometries (indicated by the
shaded region around the simulation line).}
\textcolor{blue}{Scale bars show 1~mm.  Shaded regions on measured
data represents errors estimated from standard deviation of 9 scans,
full details in SM.}}
\end{figure*}

The above analysis doesn't account for the finite size of the probe or long-range electrostatic interactions.
These contributions can be incorporated by replacing the simple capacitance, $C(h)$, with an integral over a distributed capacitance, $\mathcal{C}(h, x)$
\citep{Machleidt2009Jun, Cohen2013Jun}.
In this case, the SKPM condition becomes
\begin{equation}
    0 = \frac{d}{dh}\int_S (V_m(x) - V_S(\xi - x)) \mathcal{C}(h, \xi)\ d\xi,
\end{equation}
where $\xi$ is an integration variable. Solving this expression for $V_m$ gives
\begin{equation}
    %V(x) = \int_S \frac{u_1(\xi)}{u_0(x)} V_S(\xi)\ d\xi
    V_m(x) = \int_S P(x, \xi) V_S(\xi)\ d\xi
    \label{eq:psf_1}
\end{equation}
where we define the PSF, $P(x, \xi)$, which accounts for the terms involving the interaction
of the probe at location $x$ with different sample positions $\xi$.
Hence, a slightly more sophisticated analysis reveals that a raw SKPM measurement is the convolution of the underlying signal with
a point spread function corresponding to the finite size of the probe and
long-range interactions.
\textcolor{blue}{It is important to note that the PSF is not unique for a particular probe.
Instead, it depends on the sample-probe interaction through the distributed capacitance.
For example, this may give rise to completely different PSFs for purely metallic, purely insulating,
or mixed insulator-metal samples due to how the probe interacts with the surface or the bulk of
the material}\citep{Hudlet1995Apr}.

In practice, the situation is more complicated still, though for less obvious reasons. The measured signal depends not only on the electrostatic interactions
between the probe and the sample, but also on the feedback system that measures the
current and applies the voltage, $V$. Both the PID and lock-in amplifier require finite response times to produce stable signals. If the probe travels a significant distance
over these timescales, then information from a range of locations is incorporated into the measurement.
Additionally, faster scan speeds can introduce additional mechanical noise which
will further distort the measured signal.
Only for a stable system and when the probe is held at each location for
long time relative to the response times of the PID/lock-in do
we recover Eq.~\ref{eq:psf_1}.
Assuming these processes are linear, we can still write the measured voltage as a purely spatial convolution over a point spread function, but one that is velocity dependent; in other words,
\begin{equation}
    V_m(x) = \int P(v \,|\, x, \xi ) V_S(\xi)\ d\xi.
    \label{eq:psf_2}
\end{equation}
As the scan velocity approaches zero, $P(v\, |\, x, \xi ) \rightarrow P(x, \xi)$, and the ``fast scan'' regime (Eq.~\ref{eq:psf_2}) becomes equivalent to the ``slow scan'' regime (Eq.~\ref{eq:psf_1}).  

\subsection{Signal deconvolution}

In the previous section, we discussed the forward problem: given an underlying surface potential, $V_S(x)$, determine the resulting measurement, $V_m(x)$.  However, what is usually required is the solution to the inverse problem: given the measurement, $V_m(x)$, determine the underlying signal, $V_S(x)$ (\textit{i.e.}, recover Fig.~\ref{fig:figure1}\textbf{c} from  Fig.~\ref{fig:figure1}\textbf{b}).  We start by writing the convolution over the PSF more compactly as
\begin{equation}
    V_m = V_S \otimes P.
\end{equation}
We would like to find an estimate for the true surface voltage map, $\bar{V}_S$.
In the absence of measurement noise, and for a
non-vanishing $P$, we can
exactly recover the true
surface voltage map, $V_S$, by
\begin{equation}
    \bar{V}_S = V_m\otimes P^{-1} = (V_S \otimes P) \otimes P^{-1} = V_S,
    \label{eq:best_guess}
\end{equation}
where $\otimes P^{-1}$ denotes the deconvolution with respect to $P$.
This expression is relatively easy to calculate by taking advantage of the convolution theorem, \textit{i.e.}~
\begin{equation}
    \bar{V}_S = \mathcal{F}^{-1} \left(
        \frac{\mathcal{F}(V_m)}{\mathcal{F}(P)}\right),
        \label{eq:deconv_simple}
\end{equation}
where $\mathcal{F}$ and $\mathcal{F}^{-1}$ represent the Fourier transform and inverse Fourier transform, respectively.  

Real systems suffer from measurement noise, which renders the  equivalence of $\bar{V}_S$ and $V_S$ in Eq.~\ref{eq:best_guess} unattainable in practice.  Mathematically, independent sources of noise, $N$, enter into the equation as
$V_m = V_S\otimes P + N$, and contain high-frequency components which deconvolution can amplify. To reduce the effect of this, one approach is to simply apply
a low-pass filter to $V_m$, which can be heuristically motivated given the finite size of probes\citep{Pertl2022Dec}.  However, in order to avoid discarding higher frequency information unnecessarily,
an alternative is to use a weighted deconvolution, such as a Wiener filter\citep{Machleidt2009Jun, Cohen2013Jun}.  Mathematically, this manifests itself as a modification to Eq.~\ref{eq:deconv_simple}:
\begin{equation}
    \bar{V}_S = \mathcal{F}^{-1} \left(
        \frac{\mathcal{F}(V_m)}{\mathcal{F}(P)}\times
        \left(\frac{|\mathcal{F}(P)|^2}{|\mathcal{F}(P)|^2 + \frac{1}{\textrm{SNR}(f)}}\right)\right),
        \label{eq:wiener}
\end{equation}
where, $\textrm{SNR}(f)$ is the signal to noise ratio as a function of spatial frequency.
The SNR can either be measured (\textit{e.g.}, by calculating the power spectral density of representative samples with known properties), or in certain circumstances assumed in conjunction with a frequency-relationship for the noise (\textit{e.g.}~Brownian noise
$1/f^2$, pink noise $1/f$, \textit{etc.})

\section{Target Fabrication \& Measurement}
\label{sec:methods-short}

To experimentally determine the PSFs, we fabricate calibration
targets using \textcolor{blue}{materials} with different work functions.
We work with two different types: an edge
target, which produces a large signal but only provides information about
the PSF in one direction\citep{Claxton2008Jan, Zhang2012Jan};
and a disc target, which allows convenient measurement
of the full 2-dimensional PSF from a single 2-dimensional scan.
We utilise the edge target for measuring the slow scan PSF (Eq.~\ref{eq:psf_1})
where we assume rotational symmetry of the PSF and are interested in a
very accurate low noise estimate for the probe PSF.
We use the disc target for estimating the fast scan PSF
(Eq.~\ref{eq:psf_2}) where the PSF is highly non-axisymmetric and
our priority is on performing measurements in reasonable amounts of
time under practical conditions.

We fabricate both targets using electron beam evaporation to deposit
thin layers of different metals;
full details are provided in the Supplemental Materials (SM).
The edge target consists of a layer of platinum deposited on a
titanium coated glass slide.
The physical height of the platinum layer is small (12~nm) compared to the
sample-tip separation (60~$\text{\textmu m}$) so that its geometric influence can be ignored.
These metals are chosen for their relatively large work function
difference, which due to Fermi equalisation leads to a large
contact potential difference and correspondingly large SKPM signal ($\sim$1~V).
The disc target consists of a small, circular (400~$\text{\textmu m}$ diameter)
gold disc deposited on a silicon wafer,
creating a contact potential difference on the order of 0.5~eV.
As with the edge target, the height of the gold-titanium disc is small
(103~nm) compared to the scan height.
\textcolor{blue}{Optical microscopy images of the disc calibration
targets are included in the SM.}

We perform SKPM measurements using a commercially available device
(Biologic, M470).
\textcolor{blue}{This instrument uses a piezo to oscillate the probe,
we use an oscillation frequency of 80~Hz as we notice higher frequencies
increase acoustic noise.
The oscillation frequency sets of limit on the measurement bandwidth
and the achievable resolution at higher scan speeds, as we
discuss further in section~\ref{sec:rapid-measurement}.
We use a closed-loop type SKPM measurement, where a PID feedback controller
attempts to minimise the current in the probe.
The choice of feedback parameters are important to achieving the
best possible bandwidth: too slow a response rate will lead to a broader
PSF, while overly aggressive settings can lead to spurious oscillations in
the measured signal.
Briefly, we use the Ziegler-Nichols method\citep{Ziegler1942} to choose initial guesses for the
feedback settings and then make minor adjustments to optimise the results
(further details provided in SM).
}

When SKPM is used for work function measurements, the
signal ($\Delta\Phi$) is a relative measurement of the difference in
work function between the probe and sample.
\textcolor{blue}{To convert Kelvin probe voltages to absolute
work function values we use
a reference sample of highly ordered pyrolytic
graphite (HOPG)}\citep{Hansen2001Jun, FernandezGarrillo2018Apr},
\textcolor{blue}{additional details are given in the SM.}
The work function is then given by 
%\begin{equation}
   $ \Phi = \Delta\Phi + \Delta\Phi_{\textrm{HOPG}}$,
%\end{equation}
where $\Delta\Phi_{\textrm{HOPG}}$ is the difference between the literature value for the HOPG work function and the measured value.
To estimate the SNR for deconvolution (\emph{i.e.}, Eq.~\ref{eq:wiener}), we look at the
power spectral density of a relatively flat region and use this
to estimate a power-law relationship between SNR and frequency.

To demonstrate the effect of acquisition parameters and the
temporal dependence of acquired measurements, we perform slow scans with
step-mode acquisition and fast scans with sweep-mode acquisition.
The step-mode acquisition entails moving the probe between specific points
and holding it fixed until the
SKPM signal has stabilised (\textit{i.e.}, until we can ignore effects from the
movement and settling time of the PID/lock-in amplifier).
Sweep-mode acquisition involves moving the probe continuously across the
sample at a constant velocity, resulting in a temporally distorted signal
depending on the scan speed and feedback parameters.

 %In order to convert the relative differences to a absolute work function values, we used highly ordered pyrolytic graphite (HOPG) as a reference \cite{Salerno2018Jun}.
%Although this approach gives us an absolute value for the surface work function, the measured values can differ significantly from those reported for pure/clean samples due to, for example, adsorbed water\citep{Sugimura2002Feb}, corrosion/oxidisation, or surface contamination\citep{Yu2023Jan, Bai2023Aug, Turetta2021May, Nazarov2012}.

\section{Results \& Discussion}
\label{sec:results}

\subsection{Experimentally determining PSFs in the quasi-static limit}

\begin{figure}
\includegraphics[width=3.37in]{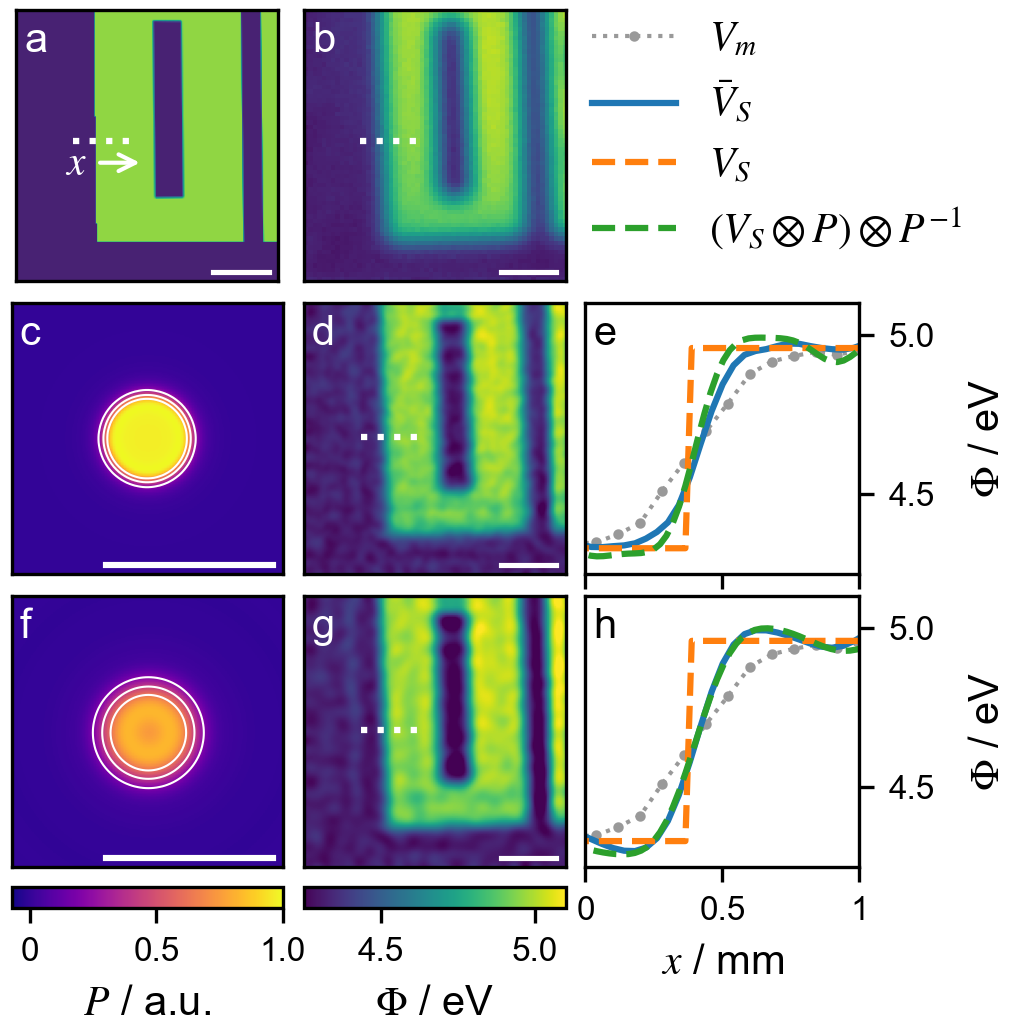}
\caption{\label{fig:figure3}
\textbf{Comparison between deconvolution with different PSFs in slow scans.}
\textbf{a}~Scanning a gold-on-silicon target~($V_S$) with sharp features 
produces \textbf{b}~a measured~($V_m$) signal with low contrast and blurred-out features.
\textbf{c--e}~Deconvolution with \textbf{c}~the simulated PSF from Fig.~\ref{fig:figure2}\textbf{d} produces
\textbf{d}~an image with higher contrast; however, as shown in \textbf{e}, a slice through
the deconvolved image ($\bar{V}_S$) differs from the ideal
scenario ($(V_S \otimes P) \otimes P^{-1}$).
\textbf{f--h}~The experimentally measured probe PSF produces a
improved estimate for the original signal.  Scale bars show 1~mm.}
\end{figure}

We now demonstrate how a PSF can be determined experimentally.
We start in the slow scan regime, \textit{i.e.}~at speeds that are small
enough such that Eq.~\ref{eq:psf_1} applies.
The main difficulty in measuring the slow scan PSF is signal strength
compared to measurement noise.
Signal strength can be improved by increasing the work function difference
between the target and surface or increasing the size of the target,
while noise can be reduced with repeated measurements and reduced scan
speed.
We have found that a good solution for the slow-scan regime is to perform a 1-dimensional scan over a sharp edge between two
materials with distinct work functions (Figure~\ref{fig:figure2}\textbf{a}).
\textcolor{blue}{Sharp edges are a common technique for estimating
the accuracy and resolution of both SKPM and KPFM systems}
\citep{McMurray2002Feb, Zerweck2005Mar, Wicinski2016Mar},
\textcolor{blue}{and the can provide enough information to estimate
the PSF for a large cylindrical probe.}
Even so, extremely slow speeds are required for probe motion to be
completely negligible.
We utilize a step size of 5~$\text{\textmu m}$,
moving at 5~$\text{\textmu m}$/s between points and dwelling at each
point for 3.5~s.
To improve SNR, we repeat and average multiple lines. In this process, a \textit{single} line scan takes approximately 45 minutes, and the ensemble takes more than seven hours. The final result is shown in Fig.~\ref{fig:figure2}\textbf{b}.

\begin{figure}
\includegraphics[width=3.37in]{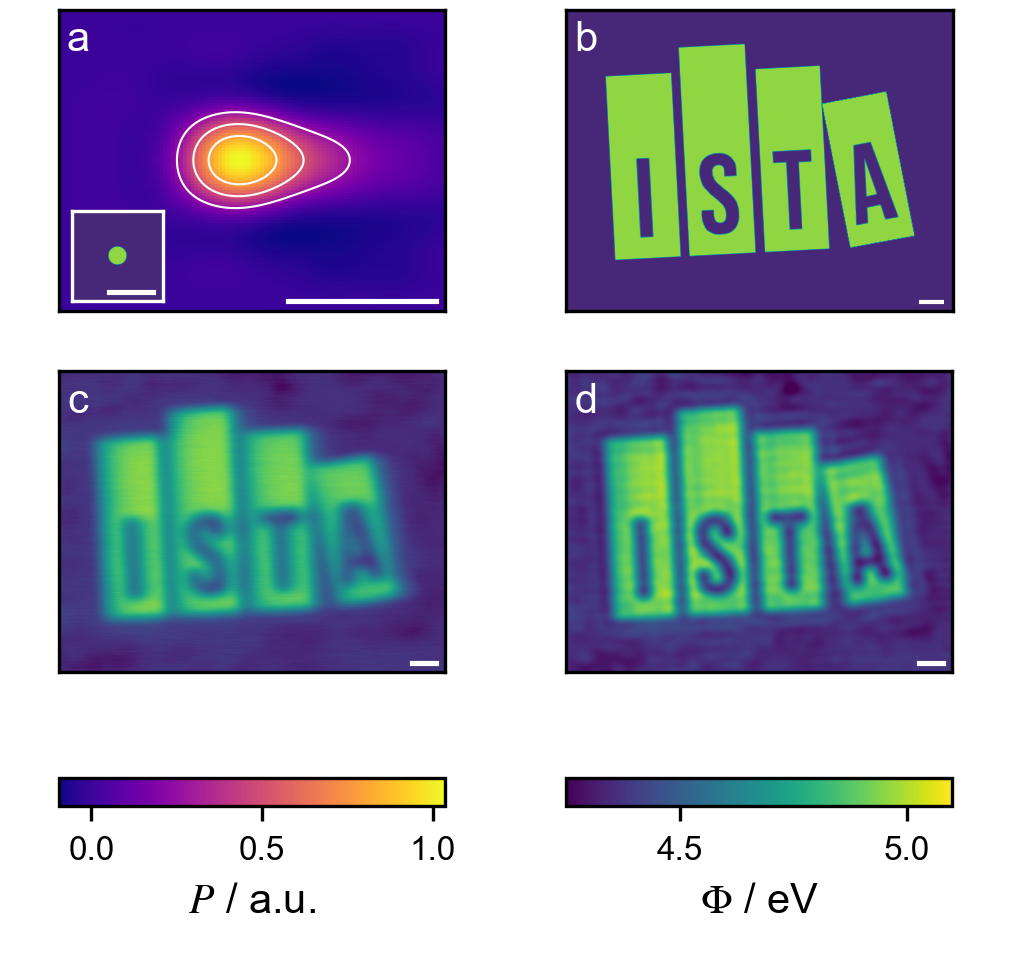}
\caption{\label{fig:figure4}
\textbf{PSF and deconvolution for a fast scan.}
\textbf{a}~Fast scans produce a significantly broadened and asymmetric
PSF, particularly along the scan direction.  \textbf{b,c}~When the same
parameters are used to scan a large target (\textbf{b}),
this produces a blurred-out image (\textbf{c}).
\textbf{d}~Deconvolution of the measured signal with the PSF
produces a significantly improved
estimate for the underlying signal.
Scale bars correspond to 1~mm.}
\end{figure}

%The ESF captures information about the tip response in one direction.
To obtain the PSF, \textcolor{blue}{we first average multiple scans of the edge to produce a low noise
estimate for the edge spread function (ESF)}.
\textcolor{blue}{We then} calculate the line spread function (LSF)
from the derivative of the line scan:
\begin{equation}
    \color{blue}
    \text{LSF}(x) = \frac{d}{dx} \text{ESF}(x).
\end{equation}
\textcolor{blue}{In} Fourier space we interpolate
the 1-dimensional transform of the line spread function to
a 2-dimensional map assuming rotational symmetry
\textcolor{blue}{and take the inverse Fourier transform to give the PSF.
Equivalently, we solve}
\begin{equation*}
    \color{blue}
    \mathcal{F}_{2D} [ \textrm{PSF} ](k_i, k_j)
        = \mathcal{F}_{1D} [ \textrm{LSF} ]\left(\sqrt{k_i^2 + k_j^2}\right) \\
\end{equation*}
\textcolor{blue}{for the PSF, where
$\mathcal{F}_{1D}$ and $\mathcal{F}_{2D}$ denote the 1-D
and 2-D Fourier transforms, and $k_i, k_j$ are the Fourier
space coordinates }(see SM for full details).
The result is shown in (Fig.~\ref{fig:figure2}\textbf{c}).
The cross section of this PSF (Fig.~\ref{fig:figure2}\textbf{e}) has a non-intuitive feature: a ring of higher intensity towards the edge of the probe.
To investigate this further and perform a sanity-check of our strategy, we perform numerical simulations for a second,
independent estimate of the PSF (Fig.~\ref{fig:figure2}\textbf{d,e}; see SM for details on simulations).
\textcolor{blue}{In addition, the PSF also has an apparent negative region at around
$\pm$0.8~mm; however, this measurement is not statistically significant.}

The simulations produce a decent estimate for the PSF, but the differences are non-negligible.
The simulated PSF has a higher amplitude, is slightly narrower, and exhibits a more subdued ring at the edge. We suspect these differences are due to physical features of the probe that the idealised simulation geometry cannot capture.
\textcolor{blue}{Our first thought was that the difference was related
to an incomplete model of the probe geometry.}
\textcolor{blue}{We included the shield in our simulations as a grounded
cylinder around the probe; however, this did not account for the
broader PSF shape.}
\textcolor{blue}{Visual} inspection of the probe reveals it is not perfectly
cylindrical \textcolor{blue}{and has a
rounded edge (see images in SM).}
\textcolor{blue}{However, simulations exploring a range of edge
radii between 5~$\text{\textmu m}$ and 80~$\text{\textmu m}$ (illustrated by the
shaded region in Fig.~\ref{fig:figure2}\textbf{e})
show that this alone is not enough to explain the broader
PSF width and scale.}
High-speed video reveals subtle horizontal vibrations in addition
to the vertical motion (see SM). \textcolor{blue}{
A possible explanation for the broader PSF is that these vibrations
result in a larger effective probe diameter.}
%A cylindrical cantilever, such as our probe, supports
%a range of vibration modes, which may cause the probe to appear
%broader.
%At higher scan speeds, and in noisy environments, }
%\begin{equation}
%    \color{blue}
%    f_0 = \frac{a_0}{2\pi} \sqrt{\frac{EI}{mL^4}}
%\end{equation}
%\textcolor{blue}{E = 340, I = $\pi r^4/4$, m = Lpir2.
%Does it behaves as a pendulum or a chain or a beam?}
%At higher speeds, or in noisy environments, this effect is likely
%to become worse (as illustrated by SM Fig.~\ref{fig:si-figure7}).}
\textcolor{blue}{The difficulty in constructing an accurate model that
fully reproduces the observations}
even in the slow-scan regime highlights the need for measuring PSFs.

With a PSF in hand, we turn our attention to reconstructing an
underlying signal source from an actual measurement.
Fig.~\ref{fig:figure3}\textbf{a} illustrates our gold-on-silicon target,
where the pattern consists a series of vertical/horizontal stripes
switching between the two materials. We scan over this target at a
slightly lower resolution (80~$\mu$m steps) and slightly
larger speed (20~$\mu$m/s with 0.6~s dwell time)
so that the scan doesn't take unreasonably long.
This produces the measured voltage map, $V_m$,
of Fig.~\ref{fig:figure3}\textbf{b}. Using Eq.~\ref{eq:wiener} and the simulated
and measured probe PSFs of Fig.~\ref{fig:figure2}, we find that we can indeed
recover estimates with improved resolution/contrast
(Fig.~\ref{fig:figure3}\textbf{c--h}).
We show in the SM that these recoveries already suffer from slightly elevated scan speed; trying to eliminate temporal information in the PSF entails compromises between speed and resolution that are difficult to balance.
As we show in the next section, the better solution is to simply incorporate the probe motion into the PSF.

\subsection{Determining PSFs for rapid measurements}
\label{sec:rapid-measurement}

Without compromising scan speed, we would like to be able to characterise large patterns of interest with high spatial resolution. As mentioned previously, when the distortions introduced by fast scanning are linear, then one can anticipate a velocity-dependent PSF, $P(\, v \, | x, \xi)$, that nonetheless connects the measured signal, $V_m$, to the underlying signal, $V_S$, via spatial convolution, as in Eq.~\ref{eq:psf_2}. In order to acquire this PSF, we straightforwardly perform a scan with the same measurement parameters as we use for the sample measurement.
While we could repeat the procedure involving the edge PSF described above,
we now no longer have to worry about scanning slowly to stay in a regime where
Eq.~\ref{eq:psf_1} is applicable.
Moreover, scanning fast creates an additional source of broadening to the PSF,
making the use of larger targets more practical.

To get a velocity-dependent scan, we now operate the SKPM in sweep (continuous) mode
(as opposed to step mode), and with a significantly higher speed
of 200~{\textmu}m/s.
\textcolor{blue}{To estimate these fast scan PSFs, we explored using 300, 400, and 500~$\mu$m diameter discs.
Ideally we would want to use a target that is much smaller than the probe diameter; however,
we found that the smallest disc produced a signal that was too weak.
For 200~$\mu$m/s scans, the 400~$\mu$m disc was sufficient, but for
the faster 1000~$\mu$m/s scans we needed to use the 500~$\mu$m diameter disc.
In order to account for the finite size of the disc, we deconvolved the measured
potential by a circular aperture and used a low-pass filter to remove high frequency noise.
Full details are provided in the SM.}

Figure~\ref{fig:figure4} shows how we put this into practice.
\textcolor{blue}{Fig.~\ref{fig:figure4}\textbf{a}~shows the PSF estimated from the scan of the disc shown
in the inset.}
As is visually apparent in the image of the PSF,
the velocity dependence blurs the image in the scan direction (left to right).
Next, we scan a large and detailed target ($\sim$1.5$\times$1.0~cm$^2$), as shown in Fig.~\ref{fig:figure4}\textbf{b}, which again consists of gold patterned onto a silicon wafer. The scan parameters are intentionally set to be the same as in the PSF of Fig.~\ref{fig:figure4}\textbf{a}.
The resulting raw voltage map is shown in Fig.~\ref{fig:figure4}\textbf{c}.
Comparing this to the slower scans of Fig.~\ref{fig:figure2}, the spatial blurring is much more apparent \textcolor{blue}{-- this is in
part due to the higher scan speed and chosen feedback parameters.}
Moreover, it is evident that this has the same left-to-right tail as the corresponding PSF. However, upon spatially deconvolving this scan with the 
corresponding PSF, the blur is reduced (Fig.~\ref{fig:figure4}\textbf{d}).
%We validate this quantitatively in Fig.~\ref{fig:figure4}\textbf{e}, where we plot the target, raw, and recovered values on top of each other, illustrating that the recovered voltages are much sharper and lie much closer to the target values.

\begin{figure}
    \centering
    \includegraphics{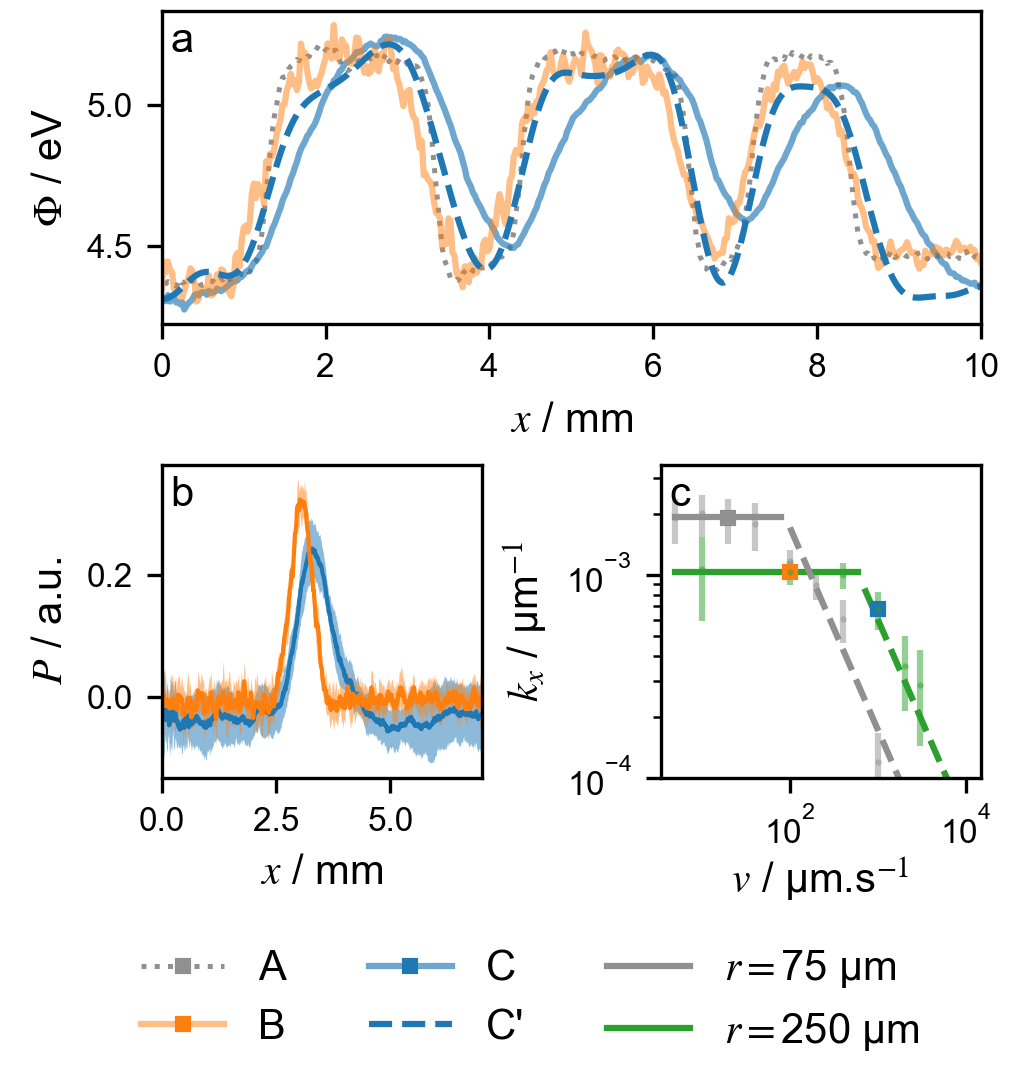}
    \caption{\color{blue}
    \textbf{Comparison between fast, deconvolved and slow scans.}
    \textbf{a}~Partial scan across the target from Fig.~\ref{fig:figure4}
    using (A)~a $r=75$~$\mu$m radius probe scanned slowly and
    averaged over 4 scans,
    and single scans with a $r=250$~$\mu$m radius
    probe at (B)~100~$\mu$m/s and (C)~1000~$\mu$m/s.
    (C')~shows the fast scan deconvolved by a PSF estimated from
    scans of a 250~$\mu$m radius disc scanned using the same
    scan parameters.
    \textbf{b}~Scans of the calibration disc with the large probe
    clearly showing the broadening due to higher scan speed.
    \textbf{c}~The effective spatial cut-off frequency, $k_x$,
    estimated for different scan speeds~($v$)
    with two different probe radii~($r$) using the same scan
    parameters as in \textbf{a}.  Squares mark the scans corresponding
    to the speeds shown in \textbf{a,b}.
    Error bars in \textbf{b} are estimated from the standard deviation
    of 10 scans.
    Error bars in~\textbf{c} are estimated form the spatial frequency
    when the signal drops bellow 10\% in frequency space.
    }
    \label{fig:figure5}
\end{figure}

{\color{blue}

It is clear from these scans that the PSF shape and resulting signal depend
on the acquisition speed; however, it is unclear how the observed
broadening depends on the probe size, the introduction of additional
noise from scanning at high speeds, the feedback system, and the scan velocity.
To explore this further, Figure~\ref{fig:figure5} shows line scans of the target
at different speeds using two different probes.
The feedback system was tuned to give the fastest possible response to a
gold-silicon work function step without introducing
too much additional noise.
At slow speeds, we see that these settings are sufficient to track the
signal given by a small probe scanned slowly over the same region.
At high speeds, we see a significant variation in estimated work function --
there is a significant lag along the scan direction.
Upon deconvolving the measured signal by a PSF measured at the same speed,
we see qualitatively that we recover a significantly improved estimate
for the potential.
This is very apparent in a line scan across a 500~$\mu$m diameter disc
(Fig.~\ref{fig:figure5}\textbf{b}), where we see the slow scan produces a
scan resembling the convolution of two similarly sized circular apertures,
while the fast scan produces a reduced signal with a wider distribution.
}

{\color{blue}
Unlike in the quasi-static case, the broadening behaviour here is dominated
by how quickly the feedback system can respond to a change in the surface
work function.
Effectively, the highest resolvable spatial frequency in the scan direction
is give by
\begin{equation}
    k_{x} = \min \left[ k_{\text{probe}},\ \frac{\omega}{v} \right],
\end{equation}
where $v$ is the scan velocity, $k_{\text{probe}}$ is the spatial resolution
limited by the probe geometry, and $\omega$ describes the measurement
cut-off frequency.
The measurement cut-off frequency depends on both the response rate of
the system (i.e., the SKPM oscillation frequency and PID update rate)
as well as the chosen feedback parameters,
which in turn depend on the signal-to-noise ratio.
As a first order approximation,
the SKPM signal scales with the probe cross-sectional area;
while the noise (ignoring the effects of measurement noise) scales with
the square-root of the current signal.
This results in a effective bandwidth of
\begin{equation}
    \omega \propto \Delta_V r^2
\end{equation}
where $\Delta_V$ is the minimum detectable signal.
This trend is illustrated by
Fig.~\ref{fig:figure5}\textbf{c}, which shows the $1/v$ fall-off at higher
scan speeds and the probe size dependence in the corner frequency
(additional measurements showing the noise at different velocities are
included in SM).
In practice, the cut-off frequency will be lower for a poorly optimised
PID and other sources of noise may be dominate at different scan speeds
or probe sizes.
}

\section{Conclusion}

In this work we demonstrate how the PSFs relevant to scanning Kelvin probe
microscopy (SKPM) can be experimentally determined.
We find that measured PSFs can differ significantly from those estimated
using simulations, especially for rapid scans where the finite response of
the feedback system and additional noise further broaden the PSFs.
We demonstrate that a practical approach for accounting for the effect of noise in
a scan is to measure the PSF using similar scan parameters
to those used during measurement acquisition.
We utilize two methods for estimating PSFs: one using a edge target, which
provides good signal strength but only gives information about the PSF in one
direction; and the second using a disc, which provides
information about the full 2-dimensional PSF from a single 2-dimensional scan.
While both methods could be used to acquire PSFs, we found that the disc
was particularly useful for faster scans, where the PSFs tend to be
non-axissymetric.
The edge method is more suited to acquiring lower noise PSFs but requires assumptions about the PSF symmetry or multiple measurements with different edge angles to estimate the 2-dimensional PSF.
%This 
%when we are more interested in the physical geometry of the probe rather
%than the is only useful if one really needs to obtain the PSF corresponding to the slow scan regime, where using a line substantially reduces scan time.

\textcolor{blue}{Our focus has been on characterising the relatively large probes
used in SKPM, however, the same procedure could be applicable to characterisation of
high speed KPFM measurements.}
While the larger probes used in SKPM allow direct measurement of the
PSF, for smaller probes \textcolor{blue}{(such as those used in KPFM),}
the required spot size and the corresponding decrease
in the signal to noise ratio makes experimental measurement of such PSFs difficult.
We explore targets involving work function differences between metals,
however other approaches such as depositing charge spots or creating a
artificial potential step\citep{Brouillard2022Apr} could further improve
the signal to noise ratio.
% \citep{Zerweck2005Mar}  -- salt thing
Although we focused on SKPM, this procedure could also be useful for
characterising and correcting for temporal effects in related
methods \citep{Checa2023Nov}.
Additional modelling of the signal acquisition pipeline could also
be useful for determining the effective measurement PSFs from
simulated PSFs, this could be particularly relevant for smaller probes
such as the nano-scale probes used in KPFM.
%\citep{Brouillard2022Apr, Zerweck2005Mar}.
The results presented here could be useful for scanning larger samples,
particularly with larger probes at faster scanning speeds while achieving
decent resolution in the reconstructed images.

\section*{Supplementary Material}
A detailed description of the experimental methods, simulations and additional
supporting figures can be found in this article's supplementary material.

\section*{Acknowledgements}
This project has received funding from the European
Research Council (ERC) under the European Union’s
Horizon 2020 research and
innovation program (Grant
agreement No. 949120). This research was supported by
the Scientific Service Units of The Institute of Science
and Technology Austria (ISTA) through resources provided by the
Miba Machine Shop, Nanofabrication
Facility, Scientific Computing Facility,
and Lab Support Facility.
The authors wish to thank Dmytro Rak and
Juan Carlos Sobarzo for letting us use their equipment.
The authors wish to thank the contributions of the whole
Waitukaitis group for useful discussions and feedback.

\section*{Author Declarations}
\subsection*{Conflict of Interest}
The authors have no conflicts to disclose.
\subsection*{CRediT Author Statement}
IL: Conceptualization; Formal analysis; Simulation; Investigation;
Writing -- Original Draft. FP: Investigation; Resources.
LS: Resources. SW: Writing -- Review \& Editing; Supervision; Funding acquisition.

\section*{Data Availability Statement}
The data that support the findings of this study are available from the corresponding author upon reasonable request.

\section*{References}

%\nocite{*}
\bibliography{main}

\begin{thebibliography}{29}%
\makeatletter
\providecommand \@ifxundefined [1]{%
 \@ifx{#1\undefined}
}%
\providecommand \@ifnum [1]{%
 \ifnum #1\expandafter \@firstoftwo
 \else \expandafter \@secondoftwo
 \fi
}%
\providecommand \@ifx [1]{%
 \ifx #1\expandafter \@firstoftwo
 \else \expandafter \@secondoftwo
 \fi
}%
\providecommand \natexlab [1]{#1}%
\providecommand \enquote  [1]{``#1''}%
\providecommand \bibnamefont  [1]{#1}%
\providecommand \bibfnamefont [1]{#1}%
\providecommand \citenamefont [1]{#1}%
\providecommand \href@noop [0]{\@secondoftwo}%
\providecommand \href [0]{\begingroup \@sanitize@url \@href}%
\providecommand \@href[1]{\@@startlink{#1}\@@href}%
\providecommand \@@href[1]{\endgroup#1\@@endlink}%
\providecommand \@sanitize@url [0]{\catcode `\\12\catcode `\$12\catcode `\&12\catcode `\#12\catcode `\^12\catcode `\_12\catcode `\%12\relax}%
\providecommand \@@startlink[1]{}%
\providecommand \@@endlink[0]{}%
\providecommand \url  [0]{\begingroup\@sanitize@url \@url }%
\providecommand \@url [1]{\endgroup\@href {#1}{\urlprefix }}%
\providecommand \urlprefix  [0]{URL }%
\providecommand \Eprint [0]{\href }%
\providecommand \doibase [0]{http://dx.doi.org/}%
\providecommand \selectlanguage [0]{\@gobble}%
\providecommand \bibinfo  [0]{\@secondoftwo}%
\providecommand \bibfield  [0]{\@secondoftwo}%
\providecommand \translation [1]{[#1]}%
\providecommand \BibitemOpen [0]{}%
\providecommand \bibitemStop [0]{}%
\providecommand \bibitemNoStop [0]{.\EOS\space}%
\providecommand \EOS [0]{\spacefactor3000\relax}%
\providecommand \BibitemShut  [1]{\csname bibitem#1\endcsname}%
\let\auto@bib@innerbib\@empty
%</preamble>
\bibitem [{\citenamefont {Zisman}(1932)}]{Zisman1932Jul}%
  \BibitemOpen
  \bibfield  {author} {\bibinfo {author} {\bibfnamefont {W.~A.}\ \bibnamefont {Zisman}},\ }\bibfield  {title} {\enquote {\bibinfo {title} {A new method of measuring contact potential differences in metals},}\ }\href {\doibase 10.1063/1.1748947} {\bibfield  {journal} {\bibinfo  {journal} {Rev. Sci. Instrum.}\ }\textbf {\bibinfo {volume} {3}},\ \bibinfo {pages} {367--370} (\bibinfo {year} {1932})}\BibitemShut {NoStop}%
\bibitem [{\citenamefont {Craig}\ and\ \citenamefont {Radeka}(1970)}]{Craig1970Feb}%
  \BibitemOpen
  \bibfield  {author} {\bibinfo {author} {\bibfnamefont {P.~P.}\ \bibnamefont {Craig}}\ and\ \bibinfo {author} {\bibfnamefont {V.}~\bibnamefont {Radeka}},\ }\bibfield  {title} {\enquote {\bibinfo {title} {{Stress Dependence of Contact Potential: The ac Kelvin Method}},}\ }\href {\doibase 10.1063/1.1684484} {\bibfield  {journal} {\bibinfo  {journal} {Rev. Sci. Instrum.}\ }\textbf {\bibinfo {volume} {41}},\ \bibinfo {pages} {258--264} (\bibinfo {year} {1970})}\BibitemShut {NoStop}%
\bibitem [{\citenamefont {Nazarov}\ and\ \citenamefont {Thierry}(2019)}]{Nazarov2019Aug}%
  \BibitemOpen
  \bibfield  {author} {\bibinfo {author} {\bibfnamefont {A.}~\bibnamefont {Nazarov}}\ and\ \bibinfo {author} {\bibfnamefont {D.}~\bibnamefont {Thierry}},\ }\bibfield  {title} {\enquote {\bibinfo {title} {{Application of Scanning Kelvin Probe in the Study of Protective Paints}},}\ }\href {\doibase 10.3389/fmats.2019.00192} {\bibfield  {journal} {\bibinfo  {journal} {Front. Mater.}\ }\textbf {\bibinfo {volume} {6}},\ \bibinfo {pages} {462587} (\bibinfo {year} {2019})}\BibitemShut {NoStop}%
\bibitem [{\citenamefont {Nazarov}, \citenamefont {Olivier},\ and\ \citenamefont {Thierry}(2012)}]{Nazarov2012}%
  \BibitemOpen
  \bibfield  {author} {\bibinfo {author} {\bibfnamefont {A.}~\bibnamefont {Nazarov}}, \bibinfo {author} {\bibfnamefont {M.-G.}\ \bibnamefont {Olivier}}, \ and\ \bibinfo {author} {\bibfnamefont {D.}~\bibnamefont {Thierry}},\ }\bibfield  {title} {\enquote {\bibinfo {title} {{SKP and FT-IR microscopy study of the paint corrosion de-adhesion from the surface of galvanized steel}},}\ }\href {\doibase https://doi.org/10.1016/j.porgcoat.2011.10.009} {\bibfield  {journal} {\bibinfo  {journal} {Progress in Organic Coatings}\ }\textbf {\bibinfo {volume} {74}},\ \bibinfo {pages} {356--364} (\bibinfo {year} {2012})},\ \bibinfo {note} {application of Electrochemical Techniques to Organic Coatings}\BibitemShut {NoStop}%
\bibitem [{\citenamefont {Baytekin}\ \emph {et~al.}(2011)\citenamefont {Baytekin}, \citenamefont {Patashinski}, \citenamefont {Branicki}, \citenamefont {Baytekin}, \citenamefont {Soh},\ and\ \citenamefont {Grzybowski}}]{Baytekin2011Jul}%
  \BibitemOpen
  \bibfield  {author} {\bibinfo {author} {\bibfnamefont {H.~T.}\ \bibnamefont {Baytekin}}, \bibinfo {author} {\bibfnamefont {A.~Z.}\ \bibnamefont {Patashinski}}, \bibinfo {author} {\bibfnamefont {M.}~\bibnamefont {Branicki}}, \bibinfo {author} {\bibfnamefont {B.}~\bibnamefont {Baytekin}}, \bibinfo {author} {\bibfnamefont {S.}~\bibnamefont {Soh}}, \ and\ \bibinfo {author} {\bibfnamefont {B.~A.}\ \bibnamefont {Grzybowski}},\ }\bibfield  {title} {\enquote {\bibinfo {title} {{The Mosaic of Surface Charge in Contact Electrification}},}\ }\href {\doibase 10.1126/science.1201512} {\bibfield  {journal} {\bibinfo  {journal} {Science}\ }\textbf {\bibinfo {volume} {333}},\ \bibinfo {pages} {308--312} (\bibinfo {year} {2011})}\BibitemShut {NoStop}%
\bibitem [{\citenamefont {Bai}\ \emph {et~al.}(2021)\citenamefont {Bai}, \citenamefont {Riet}, \citenamefont {Xu}, \citenamefont {Lacks},\ and\ \citenamefont {Wang}}]{Bai2021Jun}%
  \BibitemOpen
  \bibfield  {author} {\bibinfo {author} {\bibfnamefont {X.}~\bibnamefont {Bai}}, \bibinfo {author} {\bibfnamefont {A.}~\bibnamefont {Riet}}, \bibinfo {author} {\bibfnamefont {S.}~\bibnamefont {Xu}}, \bibinfo {author} {\bibfnamefont {D.~J.}\ \bibnamefont {Lacks}}, \ and\ \bibinfo {author} {\bibfnamefont {H.}~\bibnamefont {Wang}},\ }\bibfield  {title} {\enquote {\bibinfo {title} {{Experimental and Simulation Investigation of the Nanoscale Charge Diffusion Process on a Dielectric Surface: Effects of Relative Humidity}},}\ }\href {\doibase 10.1021/acs.jpcc.1c02272} {\bibfield  {journal} {\bibinfo  {journal} {J. Phys. Chem. C}\ }\textbf {\bibinfo {volume} {125}},\ \bibinfo {pages} {11677--11686} (\bibinfo {year} {2021})}\BibitemShut {NoStop}%
\bibitem [{\citenamefont {Hackl}, \citenamefont {Schitter},\ and\ \citenamefont {Mesquida}(2022)}]{Hackl2022Nov}%
  \BibitemOpen
  \bibfield  {author} {\bibinfo {author} {\bibfnamefont {T.}~\bibnamefont {Hackl}}, \bibinfo {author} {\bibfnamefont {G.}~\bibnamefont {Schitter}}, \ and\ \bibinfo {author} {\bibfnamefont {P.}~\bibnamefont {Mesquida}},\ }\bibfield  {title} {\enquote {\bibinfo {title} {{AC Kelvin Probe Force Microscopy Enables Charge Mapping in Water}},}\ }\href {\doibase 10.1021/acsnano.2c07121} {\bibfield  {journal} {\bibinfo  {journal} {ACS Nano}\ }\textbf {\bibinfo {volume} {16}},\ \bibinfo {pages} {17982--17990} (\bibinfo {year} {2022})}\BibitemShut {NoStop}%
\bibitem [{\citenamefont {Ørjan G.~Martinsen}\ and\ \citenamefont {Heiskanen}(2023)}]{Martinsen2023}%
  \BibitemOpen
  \bibfield  {author} {\bibinfo {author} {\bibnamefont {Ørjan G.~Martinsen}}\ and\ \bibinfo {author} {\bibfnamefont {A.}~\bibnamefont {Heiskanen}},\ }\bibfield  {title} {\enquote {\bibinfo {title} {Chapter 7 - electrodes},}\ }in\ \href {\doibase https://doi.org/10.1016/B978-0-12-819107-1.00005-4} {\emph {\bibinfo {booktitle} {Bioimpedance and Bioelectricity Basics (Fourth Edition)}}},\ \bibinfo {editor} {edited by\ \bibinfo {editor} {\bibnamefont {Ørjan G.~Martinsen}}\ and\ \bibinfo {editor} {\bibfnamefont {A.}~\bibnamefont {Heiskanen}}}\ (\bibinfo  {publisher} {Academic Press},\ \bibinfo {address} {Oxford},\ \bibinfo {year} {2023})\ \bibinfo {edition} {fourth edition}\ ed.,\ pp.\ \bibinfo {pages} {175--248}\BibitemShut {NoStop}%
\bibitem [{\citenamefont {Nonnenmacher}, \citenamefont {O{'}Boyle},\ and\ \citenamefont {Wickramasinghe}(1991)}]{Nonnenmacher1991Jun}%
  \BibitemOpen
  \bibfield  {author} {\bibinfo {author} {\bibfnamefont {M.}~\bibnamefont {Nonnenmacher}}, \bibinfo {author} {\bibfnamefont {M.~P.}\ \bibnamefont {O{'}Boyle}}, \ and\ \bibinfo {author} {\bibfnamefont {H.~K.}\ \bibnamefont {Wickramasinghe}},\ }\bibfield  {title} {\enquote {\bibinfo {title} {{Kelvin probe force microscopy}},}\ }\href {\doibase 10.1063/1.105227} {\bibfield  {journal} {\bibinfo  {journal} {Appl. Phys. Lett.}\ }\textbf {\bibinfo {volume} {58}},\ \bibinfo {pages} {2921--2923} (\bibinfo {year} {1991})}\BibitemShut {NoStop}%
\bibitem [{\citenamefont {Glatzel}, \citenamefont {Gysin},\ and\ \citenamefont {Meyer}(2022)}]{Glatzel2022Feb}%
  \BibitemOpen
  \bibfield  {author} {\bibinfo {author} {\bibfnamefont {T.}~\bibnamefont {Glatzel}}, \bibinfo {author} {\bibfnamefont {U.}~\bibnamefont {Gysin}}, \ and\ \bibinfo {author} {\bibfnamefont {E.}~\bibnamefont {Meyer}},\ }\bibfield  {title} {\enquote {\bibinfo {title} {{Kelvin probe force microscopy for material characterization}},}\ }\href {\doibase 10.1093/jmicro/dfab040} {\bibfield  {journal} {\bibinfo  {journal} {Microscopy}\ }\textbf {\bibinfo {volume} {71}},\ \bibinfo {pages} {i165--i173} (\bibinfo {year} {2022})}\BibitemShut {NoStop}%
\bibitem [{\citenamefont {Melitz}\ \emph {et~al.}(2011)\citenamefont {Melitz}, \citenamefont {Shen}, \citenamefont {Kummel},\ and\ \citenamefont {Lee}}]{Melitz2011Jan}%
  \BibitemOpen
  \bibfield  {author} {\bibinfo {author} {\bibfnamefont {W.}~\bibnamefont {Melitz}}, \bibinfo {author} {\bibfnamefont {J.}~\bibnamefont {Shen}}, \bibinfo {author} {\bibfnamefont {A.~C.}\ \bibnamefont {Kummel}}, \ and\ \bibinfo {author} {\bibfnamefont {S.}~\bibnamefont {Lee}},\ }\bibfield  {title} {\enquote {\bibinfo {title} {{Kelvin probe force microscopy and its application}},}\ }\href {\doibase 10.1016/j.surfrep.2010.10.001} {\bibfield  {journal} {\bibinfo  {journal} {Surf. Sci. Rep.}\ }\textbf {\bibinfo {volume} {66}},\ \bibinfo {pages} {1--27} (\bibinfo {year} {2011})}\BibitemShut {NoStop}%
\bibitem [{\citenamefont {Axt}\ \emph {et~al.}(2018)\citenamefont {Axt}, \citenamefont {Hermes}, \citenamefont {Bergmann}, \citenamefont {Tausendpfund},\ and\ \citenamefont {Weber}}]{Axt2018Jun}%
  \BibitemOpen
  \bibfield  {author} {\bibinfo {author} {\bibfnamefont {A.}~\bibnamefont {Axt}}, \bibinfo {author} {\bibfnamefont {I.~M.}\ \bibnamefont {Hermes}}, \bibinfo {author} {\bibfnamefont {V.~W.}\ \bibnamefont {Bergmann}}, \bibinfo {author} {\bibfnamefont {N.}~\bibnamefont {Tausendpfund}}, \ and\ \bibinfo {author} {\bibfnamefont {S.~A.~L.}\ \bibnamefont {Weber}},\ }\bibfield  {title} {\enquote {\bibinfo {title} {{Know your full potential: Quantitative Kelvin probe force microscopy on nanoscale electrical devices}},}\ }\href {\doibase 10.3762/bjnano.9.172} {\bibfield  {journal} {\bibinfo  {journal} {Beilstein J. Nanotechnol.}\ }\textbf {\bibinfo {volume} {9}},\ \bibinfo {pages} {1809--1819} (\bibinfo {year} {2018})}\BibitemShut {NoStop}%
\bibitem [{\citenamefont {Cohen}\ \emph {et~al.}(2013)\citenamefont {Cohen}, \citenamefont {Halpern}, \citenamefont {Nanayakkara}, \citenamefont {Luther}, \citenamefont {Held}, \citenamefont {Bennewitz}, \citenamefont {Boag},\ and\ \citenamefont {Rosenwaks}}]{Cohen2013Jun}%
  \BibitemOpen
  \bibfield  {author} {\bibinfo {author} {\bibfnamefont {G.}~\bibnamefont {Cohen}}, \bibinfo {author} {\bibfnamefont {E.}~\bibnamefont {Halpern}}, \bibinfo {author} {\bibfnamefont {S.~U.}\ \bibnamefont {Nanayakkara}}, \bibinfo {author} {\bibfnamefont {J.~M.}\ \bibnamefont {Luther}}, \bibinfo {author} {\bibfnamefont {C.}~\bibnamefont {Held}}, \bibinfo {author} {\bibfnamefont {R.}~\bibnamefont {Bennewitz}}, \bibinfo {author} {\bibfnamefont {A.}~\bibnamefont {Boag}}, \ and\ \bibinfo {author} {\bibfnamefont {Y.}~\bibnamefont {Rosenwaks}},\ }\bibfield  {title} {\enquote {\bibinfo {title} {{Reconstruction of surface potential from Kelvin probe force microscopy images}},}\ }\href {\doibase 10.1088/0957-4484/24/29/295702} {\bibfield  {journal} {\bibinfo  {journal} {Nanotechnology}\ }\textbf {\bibinfo {volume} {24}},\ \bibinfo {pages} {295702} (\bibinfo {year} {2013})}\BibitemShut {NoStop}%
\bibitem [{\citenamefont {Machleidt}\ \emph {et~al.}(2009)\citenamefont {Machleidt}, \citenamefont {Sparrer}, \citenamefont {Kapusi},\ and\ \citenamefont {Franke}}]{Machleidt2009Jun}%
  \BibitemOpen
  \bibfield  {author} {\bibinfo {author} {\bibfnamefont {T.}~\bibnamefont {Machleidt}}, \bibinfo {author} {\bibfnamefont {E.}~\bibnamefont {Sparrer}}, \bibinfo {author} {\bibfnamefont {D.}~\bibnamefont {Kapusi}}, \ and\ \bibinfo {author} {\bibfnamefont {K.-H.}\ \bibnamefont {Franke}},\ }\bibfield  {title} {\enquote {\bibinfo {title} {{Deconvolution of Kelvin probe force microscopy measurements{\ifmmode---\else\textemdash\fi}methodology and application}},}\ }\href {\doibase 10.1088/0957-0233/20/8/084017} {\bibfield  {journal} {\bibinfo  {journal} {Meas. Sci. Technol.}\ }\textbf {\bibinfo {volume} {20}},\ \bibinfo {pages} {084017} (\bibinfo {year} {2009})}\BibitemShut {NoStop}%
\bibitem [{\citenamefont {Ren}\ \emph {et~al.}(2023)\citenamefont {Ren}, \citenamefont {Chen}, \citenamefont {Chen}, \citenamefont {Ji},\ and\ \citenamefont {Wang}}]{Ren2023May}%
  \BibitemOpen
  \bibfield  {author} {\bibinfo {author} {\bibfnamefont {B.}~\bibnamefont {Ren}}, \bibinfo {author} {\bibfnamefont {L.}~\bibnamefont {Chen}}, \bibinfo {author} {\bibfnamefont {R.}~\bibnamefont {Chen}}, \bibinfo {author} {\bibfnamefont {R.}~\bibnamefont {Ji}}, \ and\ \bibinfo {author} {\bibfnamefont {Y.}~\bibnamefont {Wang}},\ }\bibfield  {title} {\enquote {\bibinfo {title} {{Noise Reduction of Atomic Force Microscopy Measurement Data for Fitting Verification of Chemical Mechanical Planarization Model}},}\ }\href {\doibase 10.3390/electronics12112422} {\bibfield  {journal} {\bibinfo  {journal} {Electronics}\ }\textbf {\bibinfo {volume} {12}},\ \bibinfo {pages} {2422} (\bibinfo {year} {2023})}\BibitemShut {NoStop}%
\bibitem [{\citenamefont {Checa}\ \emph {et~al.}(2023)\citenamefont {Checa}, \citenamefont {Fuhr}, \citenamefont {Sun}, \citenamefont {Vasudevan}, \citenamefont {Ziatdinov}, \citenamefont {Ivanov}, \citenamefont {Yun}, \citenamefont {Xiao}, \citenamefont {Sehirlioglu}, \citenamefont {Kim}, \citenamefont {Sharma}, \citenamefont {Kelley}, \citenamefont {Domingo}, \citenamefont {Jesse},\ and\ \citenamefont {Collins}}]{Checa2023Nov}%
  \BibitemOpen
  \bibfield  {author} {\bibinfo {author} {\bibfnamefont {M.}~\bibnamefont {Checa}}, \bibinfo {author} {\bibfnamefont {A.~S.}\ \bibnamefont {Fuhr}}, \bibinfo {author} {\bibfnamefont {C.}~\bibnamefont {Sun}}, \bibinfo {author} {\bibfnamefont {R.}~\bibnamefont {Vasudevan}}, \bibinfo {author} {\bibfnamefont {M.}~\bibnamefont {Ziatdinov}}, \bibinfo {author} {\bibfnamefont {I.}~\bibnamefont {Ivanov}}, \bibinfo {author} {\bibfnamefont {S.~J.}\ \bibnamefont {Yun}}, \bibinfo {author} {\bibfnamefont {K.}~\bibnamefont {Xiao}}, \bibinfo {author} {\bibfnamefont {A.}~\bibnamefont {Sehirlioglu}}, \bibinfo {author} {\bibfnamefont {Y.}~\bibnamefont {Kim}}, \bibinfo {author} {\bibfnamefont {P.}~\bibnamefont {Sharma}}, \bibinfo {author} {\bibfnamefont {K.~P.}\ \bibnamefont {Kelley}}, \bibinfo {author} {\bibfnamefont {N.}~\bibnamefont {Domingo}}, \bibinfo {author} {\bibfnamefont {S.}~\bibnamefont {Jesse}}, \ and\ \bibinfo {author} {\bibfnamefont {L.}~\bibnamefont {Collins}},\ }\bibfield  {title} {\enquote {\bibinfo {title}
  {{High-speed mapping of surface charge dynamics using sparse scanning Kelvin probe force microscopy}},}\ }\href {\doibase 10.1038/s41467-023-42583-x} {\bibfield  {journal} {\bibinfo  {journal} {Nat. Commun.}\ }\textbf {\bibinfo {volume} {14}},\ \bibinfo {pages} {1--12} (\bibinfo {year} {2023})}\BibitemShut {NoStop}%
\bibitem [{\citenamefont {Ziegler}\ \emph {et~al.}(2013)\citenamefont {Ziegler}, \citenamefont {Meyer}, \citenamefont {Farnham}, \citenamefont {Brune}, \citenamefont {Bertozzi},\ and\ \citenamefont {Ashby}}]{Ziegler2013Jul}%
  \BibitemOpen
  \bibfield  {author} {\bibinfo {author} {\bibfnamefont {D.}~\bibnamefont {Ziegler}}, \bibinfo {author} {\bibfnamefont {T.~R.}\ \bibnamefont {Meyer}}, \bibinfo {author} {\bibfnamefont {R.}~\bibnamefont {Farnham}}, \bibinfo {author} {\bibfnamefont {C.}~\bibnamefont {Brune}}, \bibinfo {author} {\bibfnamefont {A.~L.}\ \bibnamefont {Bertozzi}}, \ and\ \bibinfo {author} {\bibfnamefont {P.~D.}\ \bibnamefont {Ashby}},\ }\bibfield  {title} {\enquote {\bibinfo {title} {{Improved accuracy and speed in scanning probe microscopy by image reconstruction from non-gridded position sensor data}},}\ }\href {\doibase 10.1088/0957-4484/24/33/335703} {\bibfield  {journal} {\bibinfo  {journal} {Nanotechnology}\ }\textbf {\bibinfo {volume} {24}},\ \bibinfo {pages} {335703} (\bibinfo {year} {2013})}\BibitemShut {NoStop}%
\bibitem [{\citenamefont {Cole}, \citenamefont {Jinadasa},\ and\ \citenamefont {Brown}(2011)}]{Cole2011Dec}%
  \BibitemOpen
  \bibfield  {author} {\bibinfo {author} {\bibfnamefont {R.~W.}\ \bibnamefont {Cole}}, \bibinfo {author} {\bibfnamefont {T.}~\bibnamefont {Jinadasa}}, \ and\ \bibinfo {author} {\bibfnamefont {C.~M.}\ \bibnamefont {Brown}},\ }\bibfield  {title} {\enquote {\bibinfo {title} {{Measuring and interpreting point spread functions to determine confocal microscope resolution and ensure quality control}},}\ }\href {\doibase 10.1038/nprot.2011.407} {\bibfield  {journal} {\bibinfo  {journal} {Nat. Protoc.}\ }\textbf {\bibinfo {volume} {6}},\ \bibinfo {pages} {1929--1941} (\bibinfo {year} {2011})}\BibitemShut {NoStop}%
\bibitem [{\citenamefont {Hudlet}\ \emph {et~al.}(1995)\citenamefont {Hudlet}, \citenamefont {Saint~Jean}, \citenamefont {Roulet}, \citenamefont {Berger},\ and\ \citenamefont {Guthmann}}]{Hudlet1995Apr}%
  \BibitemOpen
  \bibfield  {author} {\bibinfo {author} {\bibfnamefont {S.}~\bibnamefont {Hudlet}}, \bibinfo {author} {\bibfnamefont {M.}~\bibnamefont {Saint~Jean}}, \bibinfo {author} {\bibfnamefont {B.}~\bibnamefont {Roulet}}, \bibinfo {author} {\bibfnamefont {J.}~\bibnamefont {Berger}}, \ and\ \bibinfo {author} {\bibfnamefont {C.}~\bibnamefont {Guthmann}},\ }\bibfield  {title} {\enquote {\bibinfo {title} {{Electrostatic forces between metallic tip and semiconductor surfaces}},}\ }\href {\doibase 10.1063/1.358616} {\bibfield  {journal} {\bibinfo  {journal} {J. Appl. Phys.}\ }\textbf {\bibinfo {volume} {77}},\ \bibinfo {pages} {3308--3314} (\bibinfo {year} {1995})}\BibitemShut {NoStop}%
\bibitem [{\citenamefont {Pertl}\ \emph {et~al.}(2022)\citenamefont {Pertl}, \citenamefont {Sobarzo}, \citenamefont {Shafeek}, \citenamefont {Cramer},\ and\ \citenamefont {Waitukaitis}}]{Pertl2022Dec}%
  \BibitemOpen
  \bibfield  {author} {\bibinfo {author} {\bibfnamefont {F.}~\bibnamefont {Pertl}}, \bibinfo {author} {\bibfnamefont {J.~C.}\ \bibnamefont {Sobarzo}}, \bibinfo {author} {\bibfnamefont {L.}~\bibnamefont {Shafeek}}, \bibinfo {author} {\bibfnamefont {T.}~\bibnamefont {Cramer}}, \ and\ \bibinfo {author} {\bibfnamefont {S.}~\bibnamefont {Waitukaitis}},\ }\bibfield  {title} {\enquote {\bibinfo {title} {{Quantifying nanoscale charge density features of contact-charged surfaces with an FEM/KPFM-hybrid approach}},}\ }\href {\doibase 10.1103/PhysRevMaterials.6.125605} {\bibfield  {journal} {\bibinfo  {journal} {Phys. Rev. Mater.}\ }\textbf {\bibinfo {volume} {6}},\ \bibinfo {pages} {125605} (\bibinfo {year} {2022})}\BibitemShut {NoStop}%
\bibitem [{\citenamefont {Claxton}\ and\ \citenamefont {Staunton}(2008)}]{Claxton2008Jan}%
  \BibitemOpen
  \bibfield  {author} {\bibinfo {author} {\bibfnamefont {C.~D.}\ \bibnamefont {Claxton}}\ and\ \bibinfo {author} {\bibfnamefont {R.~C.}\ \bibnamefont {Staunton}},\ }\bibfield  {title} {\enquote {\bibinfo {title} {{Measurement of the point-spread function of a noisy imaging system}},}\ }\href {\doibase 10.1364/JOSAA.25.000159} {\bibfield  {journal} {\bibinfo  {journal} {J. Opt. Soc. Am. A, JOSAA}\ }\textbf {\bibinfo {volume} {25}},\ \bibinfo {pages} {159--170} (\bibinfo {year} {2008})}\BibitemShut {NoStop}%
\bibitem [{\citenamefont {Zhang}\ \emph {et~al.}(2012)\citenamefont {Zhang}, \citenamefont {Kashti}, \citenamefont {Kella}, \citenamefont {Frank}, \citenamefont {Shaked}, \citenamefont {Ulichney}, \citenamefont {Fischer},\ and\ \citenamefont {Allebach}}]{Zhang2012Jan}%
  \BibitemOpen
  \bibfield  {author} {\bibinfo {author} {\bibfnamefont {X.}~\bibnamefont {Zhang}}, \bibinfo {author} {\bibfnamefont {T.}~\bibnamefont {Kashti}}, \bibinfo {author} {\bibfnamefont {D.}~\bibnamefont {Kella}}, \bibinfo {author} {\bibfnamefont {T.}~\bibnamefont {Frank}}, \bibinfo {author} {\bibfnamefont {D.}~\bibnamefont {Shaked}}, \bibinfo {author} {\bibfnamefont {R.}~\bibnamefont {Ulichney}}, \bibinfo {author} {\bibfnamefont {M.}~\bibnamefont {Fischer}}, \ and\ \bibinfo {author} {\bibfnamefont {J.~P.}\ \bibnamefont {Allebach}},\ }\bibfield  {title} {\enquote {\bibinfo {title} {{Measuring the modulation transfer function of image capture devices: what do the numbers really mean?}}}\ }in\ \href {\doibase 10.1117/12.912989} {\emph {\bibinfo {booktitle} {{Proceedings Volume 8293, Image Quality and System Performance IX}}}},\ Vol.\ \bibinfo {volume} {8293}\ (\bibinfo  {publisher} {SPIE},\ \bibinfo {year} {2012})\ pp.\ \bibinfo {pages} {64--74}\BibitemShut {NoStop}%
\bibitem [{\citenamefont {ZIEGLER}\ and\ \citenamefont {NICHOLS}(1942)}]{Ziegler1942}%
  \BibitemOpen
  \bibfield  {author} {\bibinfo {author} {\bibfnamefont {J.}~\bibnamefont {ZIEGLER}}\ and\ \bibinfo {author} {\bibfnamefont {N.}~\bibnamefont {NICHOLS}},\ }\bibfield  {title} {\enquote {\bibinfo {title} {Optimum settings for automatic controllers},}\ }\href@noop {} {\bibfield  {journal} {\bibinfo  {journal} {Transactions of the ASME}\ }\textbf {\bibinfo {volume} {64}},\ \bibinfo {pages} {759--768} (\bibinfo {year} {1942})}\BibitemShut {NoStop}%
\bibitem [{\citenamefont {Hansen}\ and\ \citenamefont {Hansen}(2001)}]{Hansen2001Jun}%
  \BibitemOpen
  \bibfield  {author} {\bibinfo {author} {\bibfnamefont {W.~N.}\ \bibnamefont {Hansen}}\ and\ \bibinfo {author} {\bibfnamefont {G.~J.}\ \bibnamefont {Hansen}},\ }\bibfield  {title} {\enquote {\bibinfo {title} {{Standard reference surfaces for work function measurements in air}},}\ }\href {\doibase 10.1016/S0039-6028(01)01036-6} {\bibfield  {journal} {\bibinfo  {journal} {Surf. Sci.}\ }\textbf {\bibinfo {volume} {481}},\ \bibinfo {pages} {172--184} (\bibinfo {year} {2001})}\BibitemShut {NoStop}%
\bibitem [{\citenamefont {Fern{\ifmmode\acute{a}\else\'{a}\fi}ndez~Garrillo}\ \emph {et~al.}(2018)\citenamefont {Fern{\ifmmode\acute{a}\else\'{a}\fi}ndez~Garrillo}, \citenamefont {Gr{\ifmmode\acute{e}\else\'{e}\fi}vin}, \citenamefont {Chevalier},\ and\ \citenamefont {Borowik}}]{FernandezGarrillo2018Apr}%
  \BibitemOpen
  \bibfield  {author} {\bibinfo {author} {\bibfnamefont {P.~A.}\ \bibnamefont {Fern{\ifmmode\acute{a}\else\'{a}\fi}ndez~Garrillo}}, \bibinfo {author} {\bibfnamefont {B.}~\bibnamefont {Gr{\ifmmode\acute{e}\else\'{e}\fi}vin}}, \bibinfo {author} {\bibfnamefont {N.}~\bibnamefont {Chevalier}}, \ and\ \bibinfo {author} {\bibfnamefont {{\L}.}~\bibnamefont {Borowik}},\ }\bibfield  {title} {\enquote {\bibinfo {title} {{Calibrated work function mapping by Kelvin probe force microscopy}},}\ }\href {\doibase 10.1063/1.5007619} {\bibfield  {journal} {\bibinfo  {journal} {Rev. Sci. Instrum.}\ }\textbf {\bibinfo {volume} {89}} (\bibinfo {year} {2018}),\ 10.1063/1.5007619}\BibitemShut {NoStop}%
\bibitem [{\citenamefont {McMurray}\ and\ \citenamefont {Williams}(2002)}]{McMurray2002Feb}%
  \BibitemOpen
  \bibfield  {author} {\bibinfo {author} {\bibfnamefont {H.~N.}\ \bibnamefont {McMurray}}\ and\ \bibinfo {author} {\bibfnamefont {G.}~\bibnamefont {Williams}},\ }\bibfield  {title} {\enquote {\bibinfo {title} {{Probe diameter and probe{\textendash}specimen distance dependence in the lateral resolution of a scanning Kelvin probe}},}\ }\href {\doibase 10.1063/1.1430546} {\bibfield  {journal} {\bibinfo  {journal} {J. Appl. Phys.}\ }\textbf {\bibinfo {volume} {91}},\ \bibinfo {pages} {1673--1679} (\bibinfo {year} {2002})}\BibitemShut {NoStop}%
\bibitem [{\citenamefont {Zerweck}\ \emph {et~al.}(2005)\citenamefont {Zerweck}, \citenamefont {Loppacher}, \citenamefont {Otto}, \citenamefont {Grafstr{\ifmmode\ddot{o}\else\"{o}\fi}m},\ and\ \citenamefont {Eng}}]{Zerweck2005Mar}%
  \BibitemOpen
  \bibfield  {author} {\bibinfo {author} {\bibfnamefont {U.}~\bibnamefont {Zerweck}}, \bibinfo {author} {\bibfnamefont {C.}~\bibnamefont {Loppacher}}, \bibinfo {author} {\bibfnamefont {T.}~\bibnamefont {Otto}}, \bibinfo {author} {\bibfnamefont {S.}~\bibnamefont {Grafstr{\ifmmode\ddot{o}\else\"{o}\fi}m}}, \ and\ \bibinfo {author} {\bibfnamefont {L.~M.}\ \bibnamefont {Eng}},\ }\bibfield  {title} {\enquote {\bibinfo {title} {{Accuracy and resolution limits of Kelvin probe force microscopy}},}\ }\href {\doibase 10.1103/PhysRevB.71.125424} {\bibfield  {journal} {\bibinfo  {journal} {Phys. Rev. B}\ }\textbf {\bibinfo {volume} {71}},\ \bibinfo {pages} {125424} (\bibinfo {year} {2005})}\BibitemShut {NoStop}%
\bibitem [{\citenamefont {Wicinski}, \citenamefont {Burgstaller},\ and\ \citenamefont {Hassel}(2016)}]{Wicinski2016Mar}%
  \BibitemOpen
  \bibfield  {author} {\bibinfo {author} {\bibfnamefont {M.}~\bibnamefont {Wicinski}}, \bibinfo {author} {\bibfnamefont {W.}~\bibnamefont {Burgstaller}}, \ and\ \bibinfo {author} {\bibfnamefont {A.~W.}\ \bibnamefont {Hassel}},\ }\bibfield  {title} {\enquote {\bibinfo {title} {{Lateral resolution in scanning Kelvin probe microscopy}},}\ }\href {\doibase 10.1016/j.corsci.2015.09.008} {\bibfield  {journal} {\bibinfo  {journal} {Corros. Sci.}\ }\textbf {\bibinfo {volume} {104}},\ \bibinfo {pages} {1--8} (\bibinfo {year} {2016})}\BibitemShut {NoStop}%
\bibitem [{\citenamefont {Brouillard}\ \emph {et~al.}(2022)\citenamefont {Brouillard}, \citenamefont {Bercu}, \citenamefont {Zschieschang}, \citenamefont {Simonetti}, \citenamefont {Mittapalli}, \citenamefont {Klauk},\ and\ \citenamefont {Giraudet}}]{Brouillard2022Apr}%
  \BibitemOpen
  \bibfield  {author} {\bibinfo {author} {\bibfnamefont {M.}~\bibnamefont {Brouillard}}, \bibinfo {author} {\bibfnamefont {N.}~\bibnamefont {Bercu}}, \bibinfo {author} {\bibfnamefont {U.}~\bibnamefont {Zschieschang}}, \bibinfo {author} {\bibfnamefont {O.}~\bibnamefont {Simonetti}}, \bibinfo {author} {\bibfnamefont {R.}~\bibnamefont {Mittapalli}}, \bibinfo {author} {\bibfnamefont {H.}~\bibnamefont {Klauk}}, \ and\ \bibinfo {author} {\bibfnamefont {L.}~\bibnamefont {Giraudet}},\ }\bibfield  {title} {\enquote {\bibinfo {title} {{Experimental determination of the lateral resolution of surface electric potential measurements by Kelvin probe force microscopy using biased electrodes separated by a nanoscale gap and application to thin-film transistors}},}\ }\href {\doibase 10.1039/D1NA00824B} {\bibfield  {journal} {\bibinfo  {journal} {Nanoscale Adv.}\ }\textbf {\bibinfo {volume} {4}},\ \bibinfo {pages} {2018--2028} (\bibinfo {year} {2022})}\BibitemShut {NoStop}%
\end{thebibliography}%

%\appendix

\end{document}

\maketitle

\newcommand{\degree}{$^\circ$}
\renewcommand{\thepage}{S\arabic{page}}
\renewcommand{\thesection}{S\arabic{section}}
\renewcommand{\thetable}{S\arabic{table}}
\renewcommand{\thefigure}{S\arabic{figure}}

%\tableofcontents

\section{Comparison between SKPM and KPFM}

{\color{blue}

Although similar, KPFM and SKPM operate based on
different principals: SKPM involves measuring the current in a vibrating
probe, while KPFM involves measuring the forces acting on a vibrating cantilever.
Although subtle, this difference determines the resolution and range where
these two instruments are applicable: KPFM is better suited to high resolution
small range scans, while SKPM is well suited to large area scans with high speeds.
Here we include a brief discussion about the similarities/differences
between the two techniques.  We note there are many different realisations
of the two techniques (for example, open-loop vs closed-loop SKPM) and,
while we try to keep the discussion general, the following may not apply
to all instruments/techniques.

In both SKPM and KPFM, resolution typically scales with the probe radius.
Smaller probes allow for higher resolution scans, but due to their size
they may be more sensitive to noise.
%Probe radius also determines the measurement bandwidth: smaller probes
%typically have a narrower bandwidth (i.e., a 1nm probe on a 1cm cantilever
%will be less sensitive to a feature 1m away compared to a feature
%directly underneath the probe).
SKPM measures the current induced in the probe.
In typical electrometers/amplifiers, currents as small as 1fA can be detected.
This sets a lower limit on the smallest probe size that can be used in
SKPM.  Commercial instruments typically have probes ranging down to
$\mu$m to 10s of $\mu$m in size.
In KPFM, the detected signal is related to the force on the cantilever
holding the probe, allowing much smaller (i.e., nm to $\mu$m scale) probes.

While both techniques can achieve fairly high line-scan speeds
(on the order of 100~$\mu$m/s), area scan speed is directly related to
the number of 1-dimensional line scans required to scan a 2-dimensional area.
To achieve the same resolution in both the scan direction and the direction
perpendicular to the scan direction requires $l/\lambda$ line scans, where
$\lambda$ is the spatial resolution and $l$ is the scan length.
In other words, it would take a scan using a 1~nm probe 1000 times longer to
complete compared to a 1~$\mu$m probe.
Reducing the number of scan lines when using a high resolution probe will not
necessarily result in a decrease in resolution since the probe has a finite
bandwidth.
Due to the difference in probe size, and the corresponding difference in
area scan speed, these techniques are often complementary: KPFM is able to
scan small features with high resolution, while SKPM is able to scan
very large samples at rapid speeds.

Another important difference is related to the measurement frequency
(in SKPM, this is the vibration frequency and in KPFM this is usually
related to the resonance frequency of the cantilever).
In KPFM, kHz modulation frequencies can be achieved due to the high
resonance frequencies of cantilevers.
In SKPM, the increased probe size, as well as the shielding and
electronics used to measure the current result in more massive systems
and typically lower modulation frequencies (Hz-kHz).
In both SKPM and KPFM, the time resolution is directly dependent on the
detection frequency, which is typically limited by either the
detection electronics or the cantilever frequency.
%There are a range of different KPFM modes, but in general, the
%time resolution scales with the cantilever
%frequency. %\citep{Garrett2016May, Kilpatrick2022Sep}.

Despite these differences, both KPFM and SKPM are sensitive to both
long and short range electrostatic interactions.
The measured signal is often broadened due to the finite size of the
probe and long-range electrostatic interactions as well as effects of the
feedback system and scanning speed.
When these effects are linear, they can be characterised by an appropriate
point spread function (i.e., Eq.~\ref{eq:psf_1} and Eq.~\ref{eq:psf_2}).

}

\section{Experimental Methods}

\subsection{SKPM measurement}
For all SKPM measurements we used the Biologic M470 scanning kelvin probe module.
Except where otherwise noted, all measurements used
a 500~$\mu$m diameter probe (U-SKP370/1, Biologic).
The sample was mounted on a tip-tilt-rotation stage (TTR001/M, Thorlabs) with
a custom sample holder.
For most samples, we used 100~mm diameter silicon wafers as these were easy
to handle and mount on the stage.
The wafers were electrically connected to the SKPM by attaching a piece of copper
tape (AT526, Advance Tapes) near the edge of the wafer to more easily allow
connecting the M470.
To ensure a good electrical connection between the tape and wafer, we scratched the
wafer surface with a diamond scribe and used silver paint (123-9911, RS Pro) to
electrically connect the scratched region to the tape.
We connected the platinum-titanium target using a similar method except
the silver paint could be applied directly to the platinum without
needing to first scratch the surface.
Prior to measurement, samples were levelled by making multiple capacitive
height measurements at different locations across the sample and adjusting the
stage.
The probe was then positioned at the desired height above the sample
(60~$\mu$m from the surface \textcolor{blue}{for figures 2--4, and 80~$\mu$m for figure 5}).
Key scan parameters are given in Table~\ref{tab:scan-parameters}.

\begin{table}[h]
    \centering
    \begin{tabular}{l|cccc}
         & Pt-Ti Edge & Course Scan & Fast Scan & Line Scans \\\hline
         Pixel Size ($\mu$m) & 5 & 80$\times$80 & 20$\times$40 & 20 \\
         Movement Speed ($\mu$m/s) & 5 & 20 & 200 & 20 \\
         Scan Mode & Step & Step & Sweep & Step \\
         Acquisition rate (S/s) & 8000 & 8000 & -- & 8000 \\
         Acquisition duration (S) & 20000 & 4000 & -- & 12000 \\
         Pre-delay (s) & 1 & 0.1 & -- & 0.1 \\
    \end{tabular}
    \caption{Scan parameters used throughout this work for:
    platinum-titanium (Pt-Ti) Edge (Fig.~2), Course Scan (Fig.~3), Fast Scan (Figure~4),
    and Line Scans (Figure~\ref{fig:si-figure3}).
    For step-mode scans, the dwell time is estimated from acquisition duration
    and pre-delay time.}
    \label{tab:scan-parameters}
\end{table}

\subsection{Work function calibration}
We calibrated the SKPM signal by measuring the potential using
freshly exfoliated Highly Ordered Pyrolytic Graphite (HOPG, G3389, Agar Scientific).
HOPG was freshly exfoliated using
tape (Scotch Magic Tape, Scotch Brand).
Care was taken to completely remove a whole layer of graphite.
After mounting and levelling the sample, the SKPM probe was positioned 100~$\mu$m
from the HOPG surface and the potential was measured for different system
gain parameters.
Different SKPM probes can have different calibration factors.
Table~\ref{tab:wf-calibrations} compares 2 large probes (U-SKP370/1, Biologic)
and 1 small (U-SKP-150).
We noticed that the voltage offset depends not only on the
probe, but also the electrometer gain.
The relationship between offset and gain does not appear to
be independent of probe, possibly due to the shift in phase
required to maximise the signal.
To convert the relative SKPM signal to an absolute work function difference,
we compare the measured value for HOPG with the literature value of HOPG to give
us an estimate for the work function offset $\Delta\Phi_{\text{HOPG}}$.
\textcolor{blue}{This approach gives a reasonably accurate estimate
for the absolute work function.
However, due to the variability in the literature
values for the work function of HOPG
and possible variations in experimental conditions, this
calibration procedure may not be sufficient for all applications.
Alternatives include direct measurement of the sample work function,
for instance, using ultraviolet photo-electron spectroscopy.}
The results presented in this work all used Probe 3 \textcolor{blue}{as
the large probe, and Probe 2 as the small probe.}

\begin{table}[h]
    \centering
    \begin{tabular}{l|ccc}
         & Probe 1 & Probe 2 & Probe 3 \\\hline
        Diameter ($\mu$m) & 500 & 150 & 500 \\
        LIA Phase (deg) & 136 & 131 & 133 \\[0.5em]
        Voltage (V) &  &  &  \\
        $\quad$Gain: 1	& 0.016 & -0.183	& 0.006 \\
        $\quad$Gain: 10	& 0.022	& -0.167	& 0.124 \\
        $\quad$Gain: 100	& 0.0140 & -0.17	& 0.065 \\
        $\quad$Gain: 1000	& 2.67	& 4.4	& 2.5 \\
        $\quad$Gain: 10000	& 0.723	& 2.2	& 0.52
    \end{tabular}
    \caption{Measured SKPM voltages against HOPG calibration target
        for different tip and gain parameters.}
    \label{tab:wf-calibrations}
\end{table}

\subsection{Platinum-titanium sample fabrication}
To measure the point spread function from an edge (Fig.~\ref{fig:figure2}), we created a calibration
target with platinum and titanium, since these materials have a large
work function difference.
A glass slide was cleaned using Acetone (2 minutes ulstrasonic bath) and rinsed with
isopropanol before baking at 140\degree C to for 2 minutes to remove any residual water.
The glass was then cleaned using a plasma cleaner (1~minutes, 0.5~mbar O$_2$, 240~W), and
immediately transferred to the electron beam evaporator (HV Plassy, Plassys Bestek)
and pumped to 5$\times$10$^{-7}$~mBar.
3~nm titanium was evaporated at 0.1~nm/s.
The chamber was vented and a mask was applied to half of the
titanium surface (Kepton tape).
The system was pumped again to 5$\times$10$^{-7}$~mBar,
2~nm additional titanium was deposited at 0.1~nm/s and 10~nm Pt at 0.1~nm/s.
Measurements using this sample were performed throughout the following week.
We noted that the work function values changed significantly over the following
months when this sample was reused, we suspect this is due to surface
contamination or another deterioration process.

\subsection{PSF estimation from edge}
{\color{blue}
To estimate the quasi-static PSF, we first calculated the
line spread function \textcolor{blue}{(LSF)}
from the measured edge spread function \textcolor{blue}{(ESF)},
\begin{equation*}
    \color{blue}
    \textrm{LSF}(x) = \frac{d}{dx} \textrm{ESF}(x).
\end{equation*}
In order to reduce the noise, we averaged multiple scans of
the edge and low-pass filtered the result.
We took extra care to make sure the edge was
aligned perpendicular to the scan direction by first taking
multiple scans at different positions along the edge.
The mean of each scan was subtracted to account for instrument
drift between scans and the uncertainty was estimated from the
standard deviation between scans after subtracting the mean.
The numerical derivative was then calculated using a a 4-point central
difference method and the result low-pass filtered to further reduce noise.
Errors were propagated through the low-pass and numerical difference steps
}
assuming the measurement uncertainty is significantly larger than the
additional uncertainty introduced from the numerical differentiation.
To estimate the PSF from the line spread function, we calculated the
Fourier transform of the line spread function and interpolated the
1-D Fourier transform radially into 2-D before taking the inverse
Fourier transform to give the point spread function.
\textcolor{blue}{Equivalently, we solve}
\begin{equation*}
    \color{blue}
    \mathcal{F}_{2D} [ \textrm{PSF} ](k_i, k_j)
        = \mathcal{F}_{1D} [ \textrm{LSF} ]\left(\sqrt{k_i^2 + k_j^2}\right) \\
\end{equation*}
\textcolor{blue}{for the PSF, where
$\mathcal{F}_{1D}$ and $\mathcal{F}_{2D}$ denote the 1-D
and 2-D Fourier transforms, and $k_i, k_j$ are the Fourier
space coordinates.}
Errors were propagated using the same interpolation approach, however,
we note that this may underestimate the errors in the final PSF.
{\color{blue}
To compare PSFs, we apply an area normalisation to the 2-D PSF surface.
}

\subsection{Gold lift-off lithography sample fabrication}
To fabricate the gold-on-silicon samples (used in Fig.~\ref{fig:figure3}
\& Fig.~\ref{fig:figure4})
we used a photolithography procedure, briefly outlined bellow.
New silicon wafers (p-type, Boron, $\langle$100$\rangle$, polished) were cleaned in
isopropanol for 2 minutes in an ulstrasonic bath on high power
and then rinsed briefly with additional isopropanol.
Wafers were dried using nitrogen, and heated on a hot
plate at 170\degree C for 2 minutes.
An adhesion promoter (VM652 -- a solution of a-amino propyltriethoxysilane
in an organic solvent) was spin coated on
the wafer to assist in adhesion of the lift-off
resist.
A layer of LOR5B (Micro resist technology) resist was spin
coated on the wafer (2000~rpm for 45~s).
The wafers were baked on a hotplate for 5 minutes at 170\degree C.
Two layers of positive photoresist (Microposit S1818, Micro resist technology)
were then spin coated on top of the lift-off resist (6,000~rpm for 30~s).
After each photoresist coating, the wafer was baked for
1~minute at 115\degree C.

The coated wafers were then UV exposed using a mask aligner 
(EVG 610, EV Group) with a mercury bulb light source.
Patterns for the calibration targets and work function targets
were designed using Creo Parametric and printed on 0.18mm
Polyethylene terephthalate (PET) film (JD PhotoData).
The mask was mounted on a 1.5~mm glass plate to more easily load
it into the mask aligner.
The wafer and pattern was then loaded into the mask aligner and
exposed to 150mJ/cm$^2$.
Post exposure, the wafer was baked for 1~minute at 115\degree C.

The pattern was then developed by immersion in a beaker of
MF-319 Developer (Micro resist technology) for 40 seconds and then transferred to
a beaker with de-ionised water and visually inspected.
If the sample needed additional development, the sample was transferred
back into the MF-319 beaker for 5 second intervals.
The sample was then rinsed in two baths of de-ionised water and
dried with nitrogen.

For metal deposition we used an electron beam evaporator (HV Plassy, Plassys Bestek).
We first prepared the wafer surface using oxygen plasma cleaning (240~W, 25~sccm O$_2$,
1~minute).
Immediately following, we transferred the wafer to the evaporator and pumped down to
bellow $10^{-6}$~mBar.
To improve adhesion of the gold, we first deposited 3~nm titanium (0.4~nm/s)
before depositing 100nm gold (0.6~nm/s).
The increased thickness compared to the titanium-platinum target was choosen
to improve durability during handling and future reuse of these targets.

For the lift-off step, we heated DMSO (MicroChemicals) to 50\degree C in an ultrasonic bath.
The wafer was placed in the DMSO bath for 6 minutes and ultrasonic on low power
until lift-off was complete.
Finally the wafer was rinsed in isopropanol and dried with nitrogen.

\subsection{Low cost silver calibration target}
As an alternative to using targets fabricated using lithography, we also
briefly explored using conductive silver paint as a lower cost alternative.
To make the low cost target we used a metal wire
(diameter 800~$\mu$m) dipped in conductive silver paint (123-9911, RS Pro).
By carefully contacting the end of the wire with the wafer, we were able to create
spots between about 200-400~$\mu$m (often multiple attempts
were required to produce a nice spot).
\textcolor{blue}{These low cost targets could be useful for integrating onto
existing specimens where the lithography technique is unsuitable.}

\subsection{PSF estimation from disc scans}
To estimate the PSF for fast scans, we used a 400~$\mu$m diameter
gold-on-silicon discs created using the gold lift-off procedure
described above.
The measured PSFs have a lot of high frequency noise which can
adversely affect the deconvolution.  To reduce the noise,
we filter the measured PSF.
A brief summary of the filtering procedure is outlined bellow.
\begin{enumerate}
    \item \emph{Load data from SKPM output file.}
    \item \emph{De-trend data.}
        While we attempted to level each sample before scanning, there is
        often a small tilt still remaining.
        We attempt to remove this tilt by de-trending the data based on the
        average slope of the edge values.
    \item \emph{Calculate fast Fourier transform.}  We apply most of the filtering
        in frequency space.  Before calculating the Fourier transform we
        pad the measurement with zeros to avoid introducing edge artefacts.
    \item \emph{Low pass filter.}  To remove the bulk of the high frequency noise,
        we apply a low pass filter.  The choice of cut-off frequency was
        chosen empirically based on the target signal strength, probe diameter,
        and inspection of the frequency spectrum image.
    \item \emph{Target normalisation.}
        The ideal calibration target should be much smaller than the probe diameter;
        however, smaller targets result in poorer signal-to-noise ratios.
        When using larger calibration targets, we get better signal-to-noise ratios
        but we need to account for the distortion to the PSF shape.
        To do this, we divide by the expected frequency spectrum of our
        calibration target (for circular targets, this is a Jinc function).
        For calibration targets much smaller than the probe diameter, the frequency
        space normalisation has little affect since most of the information
        in frequency space is contained within the central lobe of the Jinc.
    \item \emph{Inverse Fourier transform.}  We then calculate the inverse Fourier transform
        to give the filtered PSF.  We pad the frequency space image in order to
        smoothly upsample our PSF (this is important for low-resolution PSFs).
    \item \emph{Centring, normalisation and interpolation.}
        To make it easier to work with our PSF in other routines, we normalise
        the PSF so the maximum value is 1 (discarding sign), centre the PSF,
        and create a regular grid interpolator for our PSF.
        This makes it easier to apply our pre-calculated PSFs to new datasets,
        regardless of the pixel aspect ratio or resolution.
\end{enumerate}

\section{Microscopy images of probe and targets}

\begin{figure}
    \centering
    \includegraphics{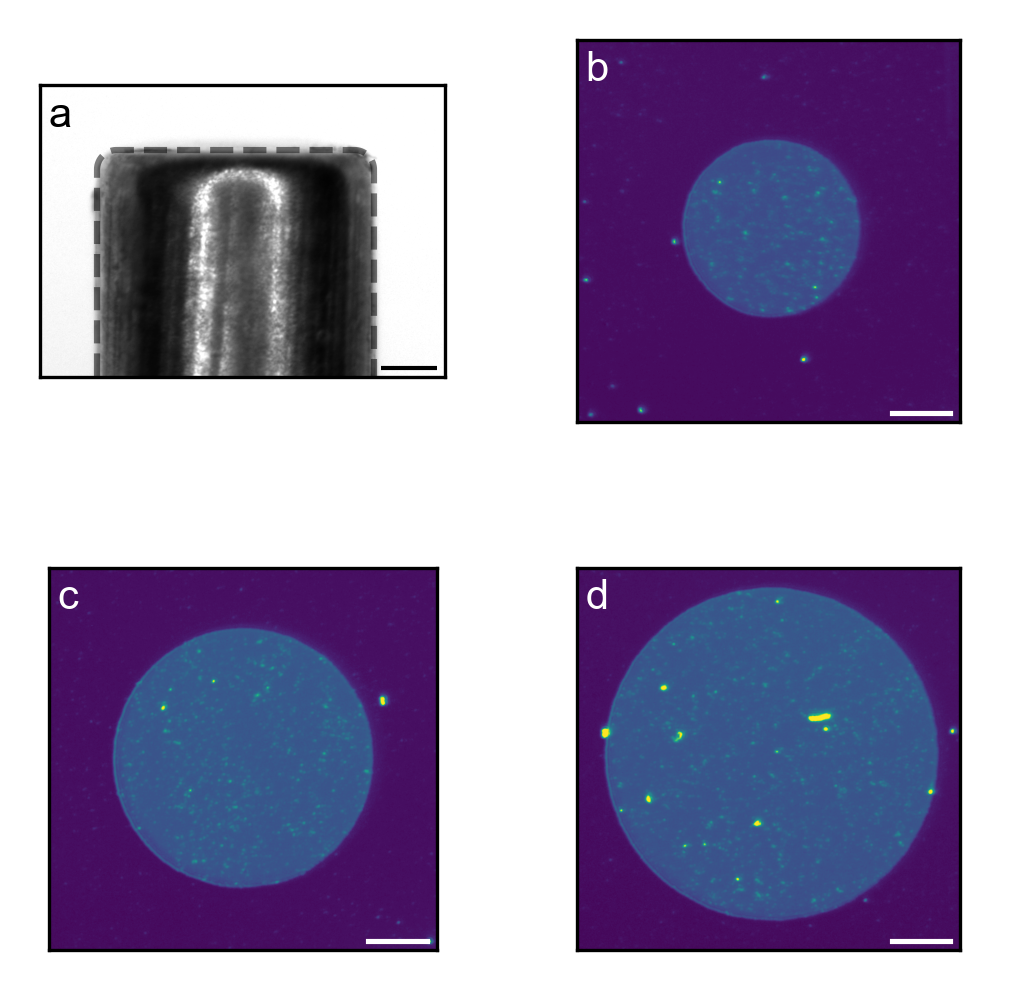}
    \caption{\color{blue}
    \textbf{Optical microscopy images of the probe and calibration targets.}
    \textbf{a}~Gray-scale image of the 500~$\mu$m probe.  The dashed outline shows the
    profile used for simulations with a 35~$\mu$m radius fillet.
    \textbf{b--d}~False-colour images of the 300, 400, and 500~$\mu$m diameter disc targets.
    Images were acquired after targets had been used extensively
    outside the clean room, explaining the accumulation of dust.
    Scale bars show 100~$\mu$m.}
    \label{fig:si-figure8}
\end{figure}

\textcolor{blue}{Figure~\ref{fig:si-figure8} shows optical
microscopy images of the circular calibration targets and the
500~$\mu$m diameter SKPM probe.
While targets were fabricated in our clean room, the SKPM is
operated outside the clean room and samples/probes acquire
dust/particulate contamination.
Compared to the probe diameter, the size of the particulate
contamination is typically small and usually only has a minor
impact on measurements, even after several months outside
our clean room.  Before use, we use compressed air to remove large
dust from the targets to avoid it potentially coming in contact
with the probe.}

\section{COMSOL simulations}
%For calculating work function PSFs, we simulated a small
%disc with fixed potential at different positions.
%We found that the disc diameter needed to be at least
%less than 10\% of the probe diameter.
%We also performed similar calculations using the Maxwell Capacitance Matrix
%and found our results to agree.

\begin{figure}
    \centering
    \includegraphics[width=3.4in]{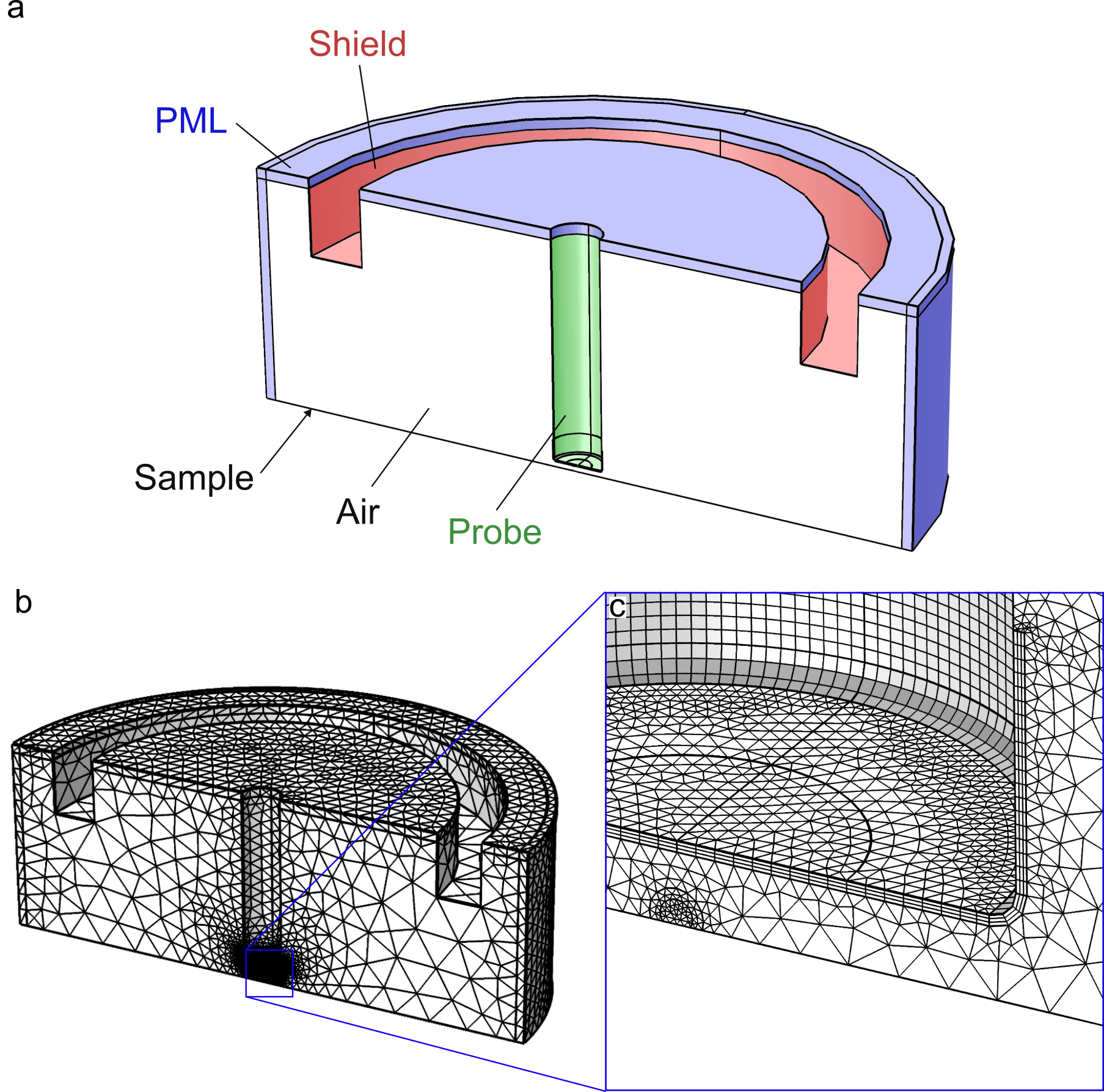}
    \caption{\textbf{a} The COMSOL model and \textbf{b,c} mesh used for PSF simulations.
    The model shows the different regions used in the COMSOL simulation including
    the perfectly matched boundary layer (PML), the conductive layers for the
    Shield, Probe and Sample, and the air volume being simulated.}
    \label{fig:si-figure2}
\end{figure}

To explore the effect of tip geometry on the probe PSF, we built a
COMSOL (5.3a) model, shown in Figure~\ref{fig:si-figure2}.
The model is mirror symmetric and includes the sample surface,
a portion of the probe and shield, and a surrounding perfectly matched
layer (PML).
Additional mesh refinement was included around the point defect and around
the tip of the probe.
The mesh size was chosen in order to produce a smooth surface charge
density on the surface and probe.
Model parameters are given in Table~\ref{tab:comsol}.

\begin{table}
    \centering
    \begin{tabular}{lr | lr}
        Parameter & Value & Parameter & Value \\\hline
        Tip radius & 0.25 mm &
        Shield outer radius & 2.79 mm \\
        PML thickness & 0.1 mm &
        Shield inner radius & 2.24 mm \\
        Tip height & 60 $\mu$m &
        Tip displacement & $\pm$15 $\mu$m \\
        Tip rounding & \textcolor{blue}{35} $\mu$m &
        Target radius & 12 $\mu$m \\
        Far-field radius & 3.35 mm &
        Shield height & 1.56 mm \\
        Far-field height & 2.34 mm & & \\
    \end{tabular}
    \caption{Key parameters used in COMSOL simulation.}
    \label{tab:comsol}
\end{table}

To calculate the PSF using this model, we calculated the surface charge
on the probe for two different situations: when 1~V was applied to
the entire sample; and when 0~V is applied to the sample except for a
small disc with non-zero voltage.
The probe is positioned at $a = \pm$15~$\mu$m from its equilibrium position
above the sample.
The dimensionless PSF signal is estimated with
\begin{equation}
    \textrm{PSF} \propto \frac{Q(a_+, x) - Q(a_-, x)}{Q(a_+) - Q(a_-)}
\end{equation}
where $Q(a_\pm, x)$ denotes the surface charge on the tip measured with the
disc at position $x$, and $Q(a_\pm)$ is for a uniform sample potential.

\begin{figure}
    \centering
    \includegraphics[width=3.4in]{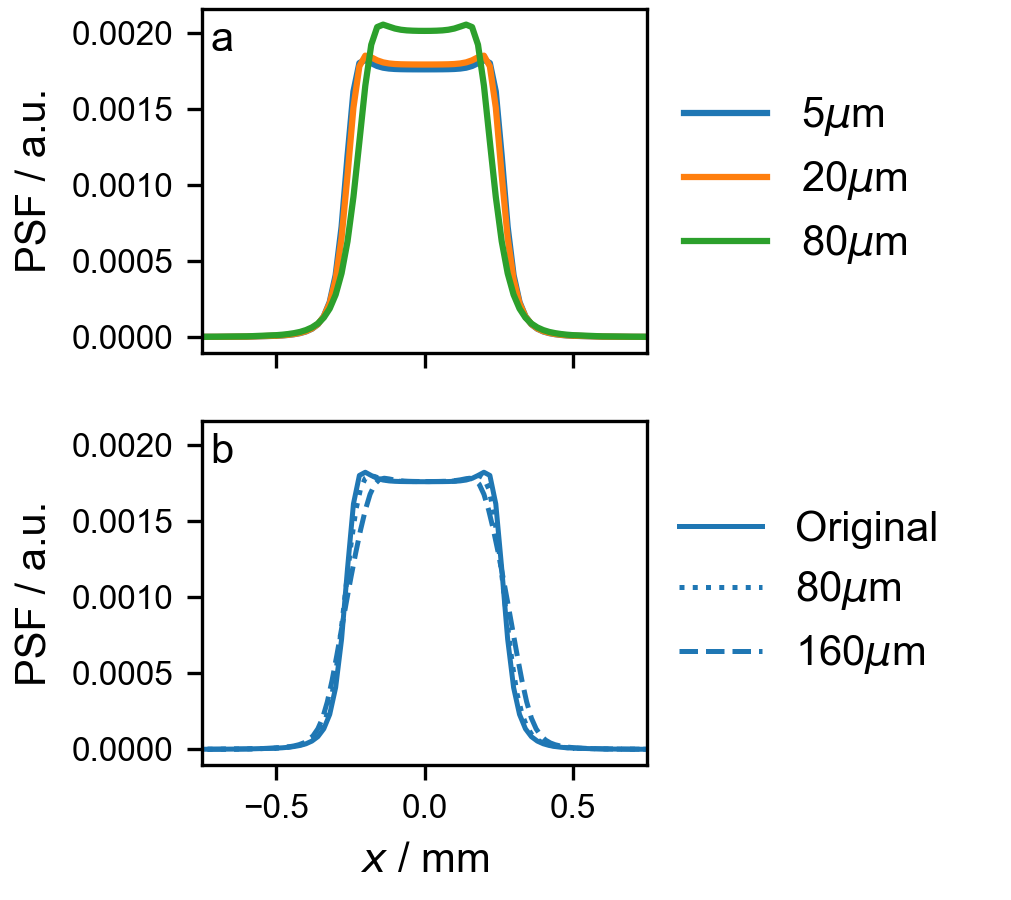}
    \caption{\textbf{a}~\textcolor{blue}{Simulated} PSFs with different tip rounding.
    \textbf{b}~Low-pass filtered PSFs \textcolor{blue}{calculated} by convolving a circular
    aperture with diameters $80~\mu$m and $160~\mu$m with the
    5~$\mu$m simulated PSF.}
    \label{fig:si-figure1}
\end{figure}

Rounding of the tip has a \textcolor{blue}{large} effect on the PSF shape.
Figure~\ref{fig:si-figure1}\textbf{a} \textcolor{blue}{compares} tips of the same diameter
but with different radius of curvature around the bottom edge of the
cylindrical shape.
Higher radius of curvature produces a more narrow PSF, but all PSFs still
have the same flattop shape with peaks near the edge.
\textcolor{blue}{Other factors are likely to have a large effect on the PSF
shape, for instance, size, uniformity of the probe, oscillation amplitude,
and height above the sample.
More extensive simulations may produce a improved estimate for the PSF shape.}

To explore the effect of a band-limited measurement of the PSF (such as
for too course resolution or too high scan speeds), we also looked at
low-pass filtering the 5~$\mu$m PSF (Figure~\ref{fig:figure1}\textbf{b}).
The low-pass filter removes the edge features.
This is not the behaviour we see in the experimentally measured PSF in
Figure~\ref{fig:figure2}\textbf{e}.
Instead, this leads us to believe that the difference between the simulated
and measured PSFs is related to our model for the tip geometry (the
physical probe is not as uniform and visible scratches can be seen in the
tip, possibly caused by use or manufacturing defects).

\section{Fast edge scan PSF estimation}
\begin{figure}
    \centering
    \includegraphics[width=3.4in]{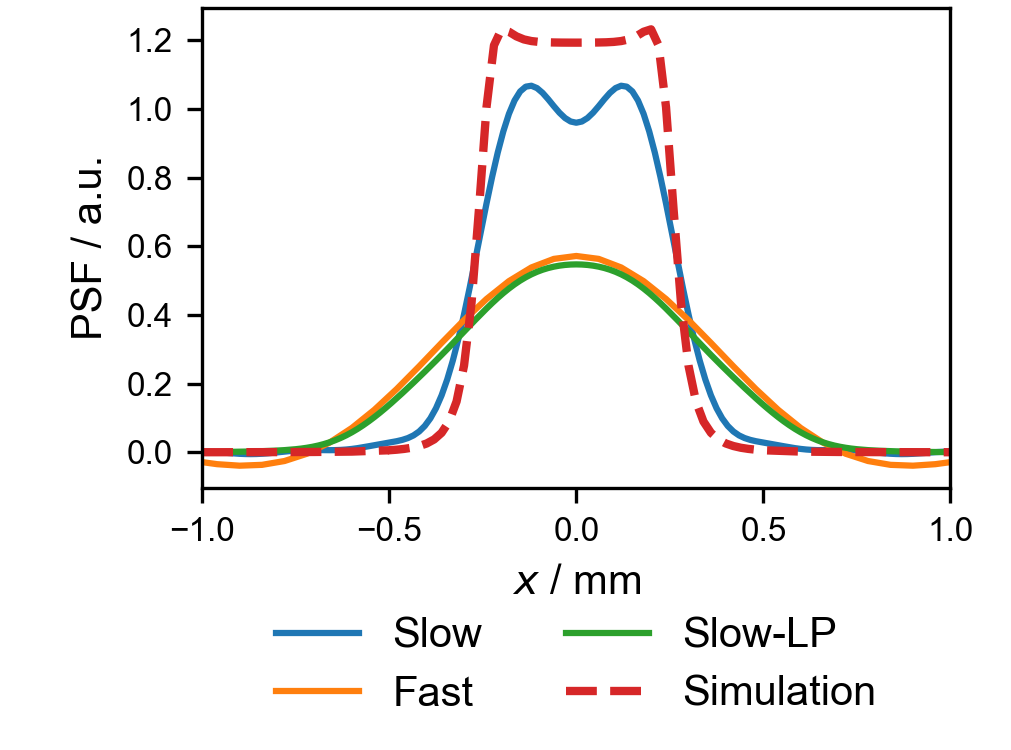}
    \caption{PSFs constructed from an edge scan using a slow (5$\mu$m/s) and
    fast (100$\mu$m/s) scan speeds.  The simulation and slow scan correspond to the
    results presented in Figure~\ref{fig:figure2}.  Slow-LP shows the fast scan
    with a 800~$\mu$m low pass filter.}
    \label{fig:si-figure4}
\end{figure}

The edge target can also be used for PSF estimation for fast scans as long as
the scan is not so fast that the PSF becomes significantly asymmetric.
Figure~\ref{fig:si-figure4} demonstrates how the PSF shape changes when the
scan speed is increased from 5~$\mu$m/s to 100~$\mu$m/s.
This scan speed still produces a very symmetric PSF, however the PSF is
significantly broadened (equivalent to applying a 800~$\mu$m low-pass filter).
The broadening means that the effective resolution is much lower, and using
a larger calibration target for these higher speed scans
(such as a 400~$\mu$m gold disc) should not significantly affect the estimate
for the PSF
While there may still be some effect from using a 400$\mu$m disc as a calibration
target, our pre-processing step for fitting the PSF and deconvolving by
the disc diameter should reduce the effect in the overall measurement.

\section{Disc target size}

%Figure~\ref{fig:figure3}\textbf{i--k} used a PSF measured with a 2-D scan
%of a 400~$\mu$m diameter gold disc.
%In order to verify that the measured signal is not simply a consequence of
%using a large calibration target, we repeated the measurement with the
%titanium-platinum edge target.
%This result is shown in Figure~\ref{fig:si-figure4}.
%By low-pass filtering the PSf acquired using the slow scan speed, we see
%that the effective target diameter required to produce a similarly shaped
%signal to the fast PSF is about 802~$\mu$m (almost double the diameter of
%our calibration target).

\begin{figure}
    \centering
    \includegraphics[width=3.4in]{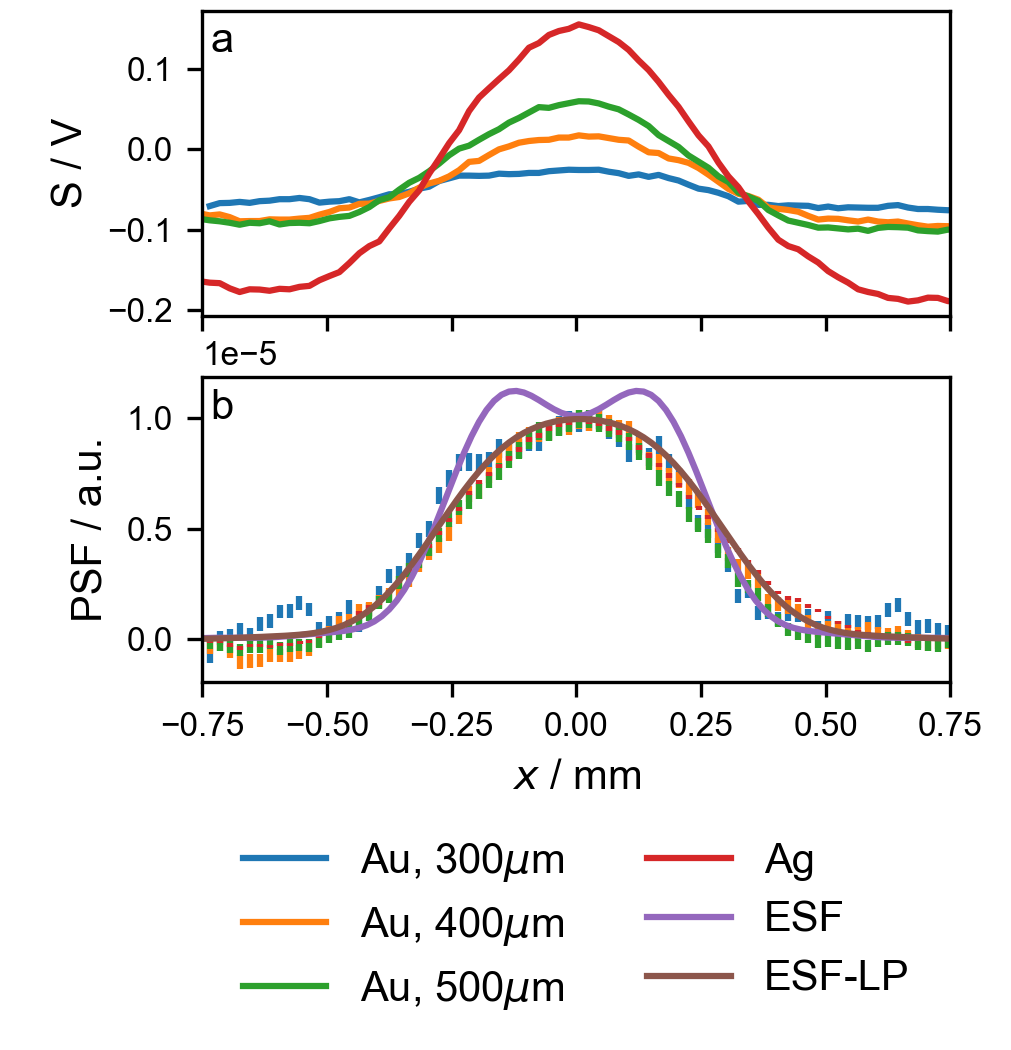}
    \caption{\textbf{a}~Line scans collected at moderate speeds of gold (Au) and
    silver (Ag) disc targets.  \textbf{b}~Estimated PSFs using the scans from \textbf{a}, with a comparison to the PSF estimated from the ESF and the PSF with a 300~$\mu$m
    low-pass filter.}
    \label{fig:si-figure3}
\end{figure}

To explore the effect of disc target size on PSF estimation, we also measured
different disc with different diameters.
The scans are shown in Figure~\ref{fig:si-figure3}.
A slightly slower scan speed was used for these scans (helping to resolve
the smaller target).
All three targets produced a similar shape, although there is still a
tiny effect of the target diameter.
As expected, the scan of the smaller disc is significantly more noisy.
These \textcolor{blue}{estimates} are all much smaller than the slow
\textcolor{blue}{measurement} PSF shown in
Figure~\ref{fig:si-figure4}, suggesting that all the targets are small enough
for the fast scan speeds used in Figure~\ref{fig:figure4}.

Figure~\ref{fig:si-figure3} also shows a line scan of a silver calibration target.
Although this target is visibly not as circular or uniform as the gold targets, 
these defects are small compared to the effective measurement PSF, and the
results are comparable to the gold targets.

\section{PSF measurement resolution}

\begin{figure}
    \centering
    \includegraphics[width=\textwidth]{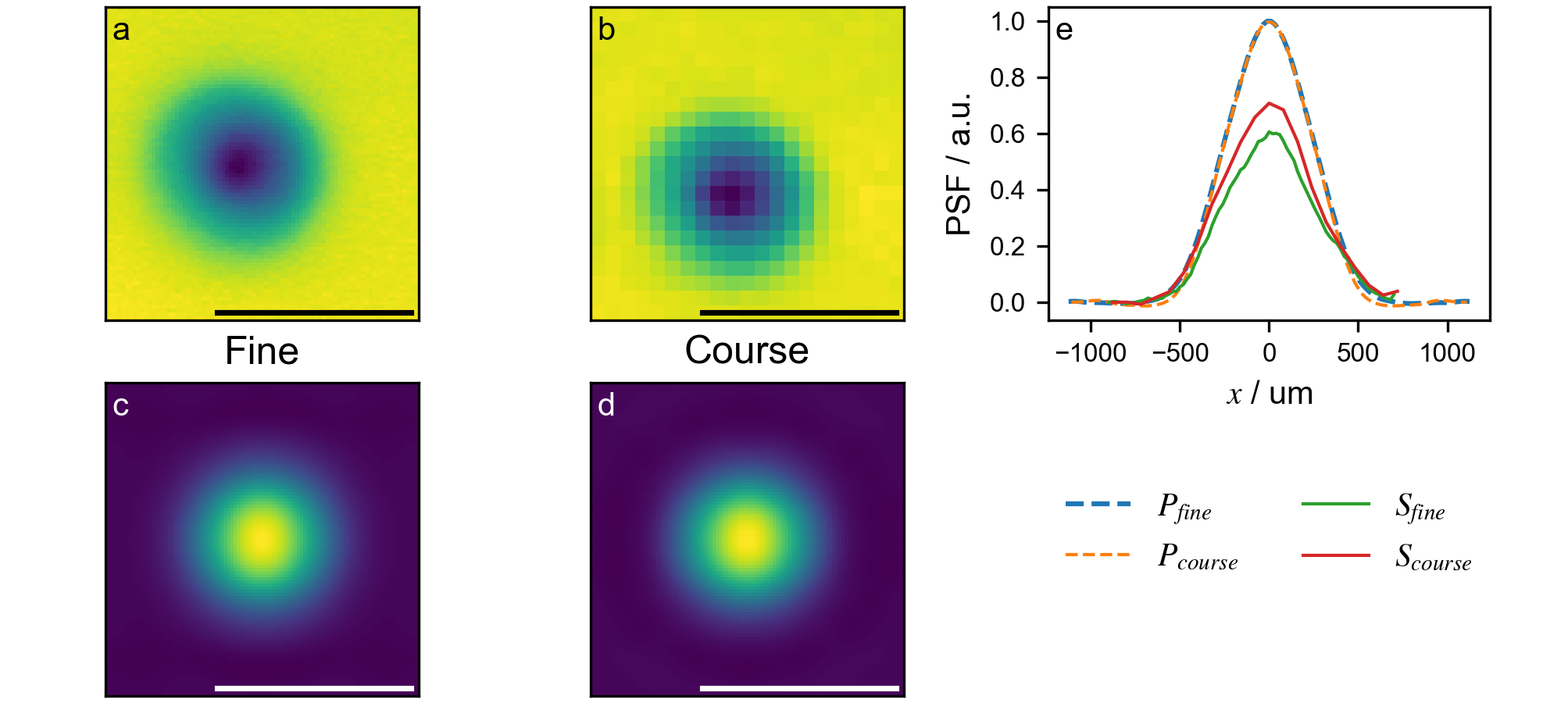}
    \caption{\textbf{a,b}~Fine and course scans of a small disc-like target.
    \textbf{c,d}~Estimated PSFs from the scan.
    \textbf{e}~Slices through centre of scans comparing profiles
    for PSFs, $P$, and measured signal, $S$.}
    \label{fig:si-figure5}
\end{figure}

Figure~\ref{fig:si-figure5} compares scan resolution and the estimated PSF shape.
While there are some variations between resolution and additional
error introduced between the
lower resolution scan, the filtering applied for estimating the PSF from these
scans seems to produce reasonable/repeatable results.

\section{Measurement of tip vibration}
\begin{figure}
    \centering
    \includegraphics[width=\textwidth]{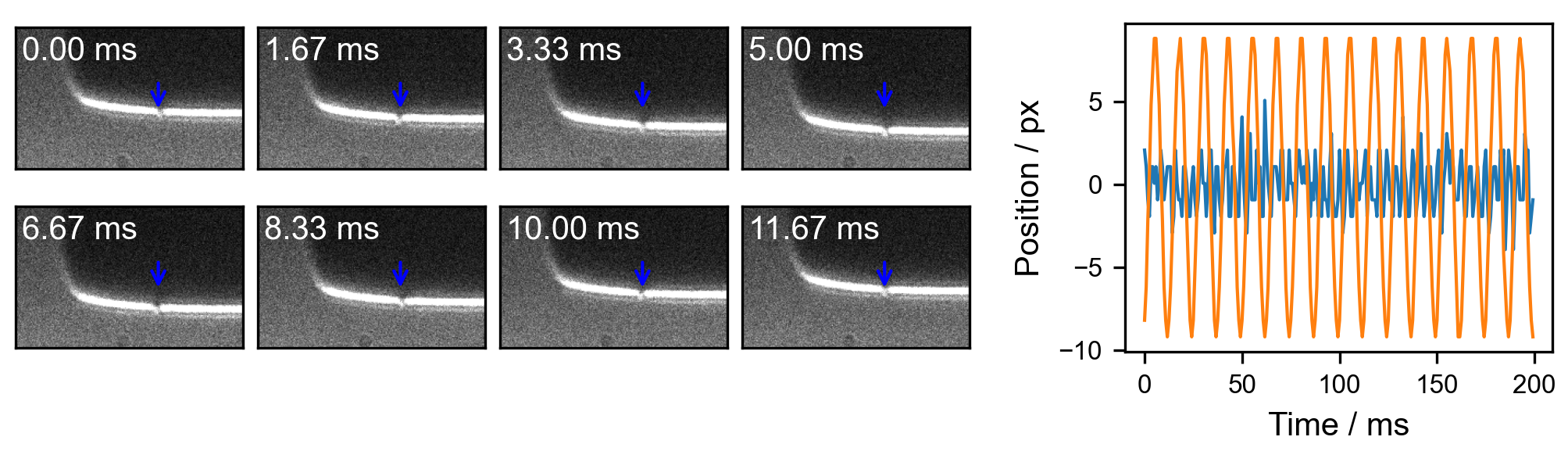}
    \caption{Still images and measured tip trajectory illustrating the motion
    of the tip.
    The arrow is meant as a visual guide, indicating a fixed reference
    spot in each image.}
    \label{fig:si-figure6}
\end{figure}

Using a fast camera (1200~frames per second), we recorded the motion of the
SKPM tip.
Stills and the estimated trajectory are shown in Figure~\ref{fig:si-figure6}.
The tip is vibrated at 80~Hz, and we can clearly see the axial motion of
the tip on the camera.  We also observe a fair amount or lateral motion,
which will likely lead to an effective larger diameter tip.
This video was recorded with the probe set to track a fixed position.
We expect that the lateral vibrations would increase when the tip is
in motion.

\section{Configuring the instrument for sweep scans}%
{\color{blue}%
Here we briefly outline the instrument configuration and tuning of the
proportional-integral-differential (PID) controller for sweep scans.
We find this procedure is suitable for the Biologic M470 SKPM, however
it may need some adaptation for other instruments.

\begin{enumerate}
    \item \textbf{Configure probe vibration}
        These settings largely depend on the chosen probe.  We choose
        settings to avoid acoustic noise.
        Higher frequencies may increase bandwidth or improve signal-to-noise.
        The instrument conditioning time
        should be set to allow any transient vibrations to be damped out
        before starting the measurement.
    \item \textbf{Configure movement settings}
        \begin{enumerate}
            \item Choose desired sweep velocity.
            \item Set the overscan.  The software recommends this be at
                least 10\% of the sweep velocity.  Higher values are better as
                they allow any transient behaviour to be damped out before the
                measurement acquisition.  In practice, this parameter is usually
                limited by the length of the sample.
            \item Set line delay time.  This is the time before starting the line scan.  This allows the PID controller to stabilise after returning.  If a high return velocity is chosen, it may be helpful to increase the line delay time.
            \item Set output time constant.  This parameter is related to the
            required voltage resolution and will directly impact achievable
            measurement resolution/bandwidth in the scan direction.
            \item Set analog-to-digital converter sampling rate,
            PID loop rate and number of samples to average. 
            The manual states that a minimum of about 20~ms is required for the device to perform calculations and respond to communications.  The software allows the PID loop rate up to 100Hz (i.e., 10ms), which still seems to produce reasonable results but the instrument seems to take longer to stop when a scan is aborted.  This parameter will also
            limit the effective resolution in fast scans.
        \end{enumerate}
        \item \textbf{Setup acquisition rates}
        \begin{enumerate}
            \item Acquisition rate can be chosen much higher than the PID update rate.  Oversampling can help reduce noise in the SKPM voltage measurement, but does not increase resolution.
            Note: if the (over-)sampling rate is not a factor of the
            PID update rate,
            this will result in a different number of points for each sample.
            \item Total number of points in scan is limited by micro-controller memory and care should be taken not to exceed it.
            \item The system will attempt to adjust the scan distance to a multiple of the step distance -- it doesn't always succeed, so its best to check this manually.
        \end{enumerate}
        \item \textbf{Determine PID starting values}
        \begin{enumerate}
            \item It is helpful to do a couple of line scans to determine these values.  Do at least two consecutive line scans.  If each line scan starts at a different value, you may need to increase the pre-line delay and/or overscan distance.
            \item Position the probe above a feature, such as an edge.
            Set the I and D term to 0.  Set a low P value and gradually increase the P value such that a scan from this position causes the signal to decay fast and just begin to oscillate.  The oscillations should be smaller than the typical feature height in the scan.
            \item There may be multiple oscillation sources, high frequency oscillations are likely from the stepper motors -- it may be better to tune the sampling rate to avoid these features.
            \item From the critical P value ($P_c$) when oscillations emerge,
            and the oscillation period ($\tau$, in seconds),
            determine suitable starting values for the I or I/D gains
            \begin{itemize}
                \item For a P controller: reduce the P value to minimise oscillations.  Approximately $P = 0.5P_c$.
                \item For a PI controller: $P = 0.45 P_c$,
                $I = \frac{1.2P}{G\tau}$,
                \item For PID: $P = 0.6P_c$,
                $I = \frac{2P}{G\tau}$,
                $D = \frac{P\tau}{8G}$.
            \end{itemize}
            where $G$ is the amplifier gain.
            At high scan speeds, the derivative term may
            amplify noise.  It may be better to use settings for a PI controller.
            If significant overshoot is observed, increasing the derivative
            factor may improve results.
        \end{enumerate}
        \item \textbf{Check performance} It is helpful to check the
        performance of the PID settings by performing a few line scans and
        inspecting the results.  If necessary, adjust the PID settings.
\end{enumerate}
}

\section{Effect of scan speed on noise}
\begin{figure}
    \centering
    \includegraphics[width=\textwidth]{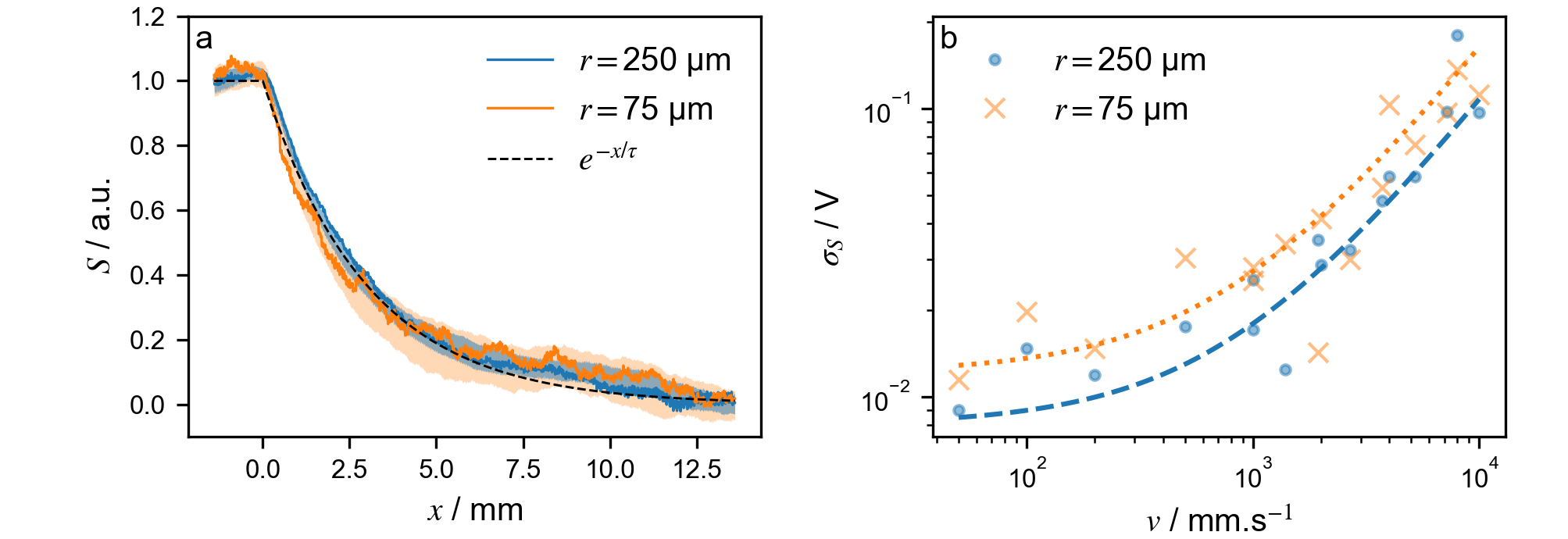}
    \caption{\color{blue}\textbf{Effect of noise on large and small probes
    for comparable response rates.}
    \textbf{a}~Smaller probes require higher gain, resulting in
    an amplification of the noise.
    Shaded regions show the standard deviation of
    10 scans with the corresponding tips.  The dashed line
    shows the exponential decay typical of an over-damped
    step response.
    \textbf{b}~Comparison of the noise as a function of
    scan velocity for different sized probes.  Each point
    represents the standard deviation of 10 scans.
    Dashed lines show the expected behaviour.}
    \label{fig:si-figure7}
\end{figure}

{\color{blue}
For high scan speeds we prefer to use larger diameter
probes, since smaller probes do not necessarily result in
improved resolution once we consider the increase in noise.
This effect can be easily observed by configuring
the PID controller so that the system is over-damped and
comparing scans of a sharp edge/feature, as shown in
Figure~\ref{fig:si-figure7}.
In general, we see a larger distribution between measurements with
smaller probes.
However, at some scan speeds we observe lower variation between scans,
possibly due to the shift in stepper motor frequency/noise relative to the
SKPM measurement frequency.
}

\section{Simulation of large ISTA target}
\begin{figure}
    \centering
    \includegraphics{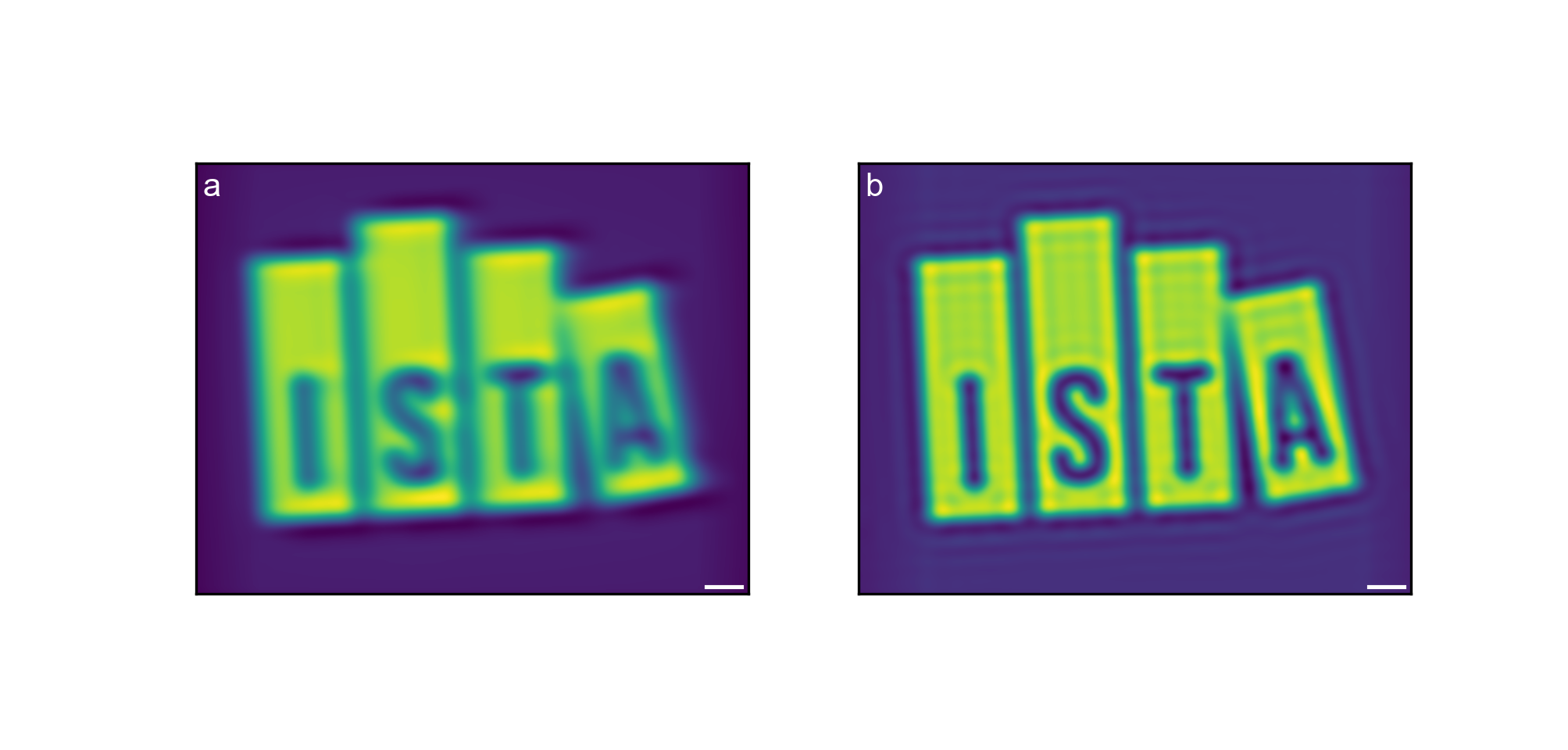}
    \caption{\color{blue}
    \textbf{Simulation of the large ISTA target.}
    \textbf{a}~Convolution of the target from Fig.~\ref{fig:figure4}a,
    with the experimentally measured PSF.
    \textbf{b}~The recovered estimate for the target after deconvolution.
    Scale bars show 1~mm.}
    \label{fig:si-figure9}
\end{figure}

\textcolor{blue}{Figure~\ref{fig:si-figure9} shows a simulation of
measurement in Fig.~\ref{fig:figure4} and the deconvolution.
The images are qualitatively very similar to the experimental measurements.}

